\documentclass{article}
\usepackage{emulateapj,apjfonts,psfig}

\begin{document}

\lefthead{Blue Spheroidal Galaxies}
\righthead{Im et al.}
\submitted{Accepted for publication in the Astronomical Journal}

\title{Are There Blue, Massive E/S0s at $z < 1$ ?: 
 Kinematics of Blue Spheroidal Galaxy Candidates}

\author{Myungshin Im\altaffilmark{1,5},
 S. M. Faber\altaffilmark{1}, 
 Karl Gebhardt\altaffilmark{2}, 
 David C. Koo\altaffilmark{1}, 
 Andrew C. Phillips\altaffilmark{1}, 
 Ricardo P. Schiavon\altaffilmark{1},
 Luc Simard\altaffilmark{3},
 \& Christopher N. A. Willmer\altaffilmark{1,4}}

\altaffiltext{1}{UCO/Lick Observatory,
 Department of Astronomy \& Astrophysics, University of
 California, Santa Cruz, CA 95064}
\altaffiltext{2}{Department of Astronomy, University of Texas, Austin}
\altaffiltext{3}{Steward Observatory, University of Arizona, 933 North
 Cherry Avenue, Tucson, AZ 85721}
\altaffiltext{4}{On leave from Observat\'orio Nacional, Rua General Jos\'e Cristino
 77, 20921-030 S\~ao Cristov\'ao, RJ, Brazil}
\altaffiltext{5}{myung@ucolick.org}

\begin{abstract} 
  Several recent studies find that 10 -- 50\% of morphologically
 selected field early-type galaxies at redshifts $z \lesssim 1$ have 
 blue colors indicative of recent star formation.
  Such ``blue spheroids''  might be
 massive early-type galaxies with active star formation,
 perhaps induced by recent merger events. 
  Alternatively, they could be starbursting, low-mass spheroids.
   To distinguish between these two choices, 
   we have selected 10 ``Blue Spheroid Candidates'' \footnote{
 Throughout this paper, ``spheroids'' are defined as featureless galaxies 
 which include Es, S0s, and dEs, while bulges of spiral galaxies
 are not included.} (hereafter, BSCs) 
 from a quantitatively
 selected E/S0 sample to study their properties,  
 including kinematics from Keck spectra obtained as part of
 the DEEP Groth Strip Survey (GSS).
  Most BSCs (70\%) turn out to belong to two 
 broad categories, while the remaining objects are likely to be 
 misclassified objects. Type-1 BSCs
 have underlying red stellar
 components with bluer inner components.  Type-2 BSCs
 do not show an obvious sign of the underlying red stellar component, 
 and their overall colors are quite blue ($(U-B)_{rest} < 0$).
  Both Type-1 and Type-2 BSCs 
 have internal velocity dispersions measured from emission lines 
 $\sigma \lesssim 80$ km~sec$^{-1}$ and estimated dynamical masses of only 
 a few $\times 10^{10}M_{\sun}$ or less.
  For Type-1 BSCs,  we estimate 
 $\sigma$ of the red component using the fundamental plane 
 relation of distant field absorption-line galaxies
 and find that these $\sigma$ estimates 
 are similar to the $\sigma$ measured
 from  emission lines. 
  Overall, we conclude that our Type-1 and Type-2  BSCs
 are more likely to be star-forming  low mass spheroids 
 than star-forming, massive, early-type galaxies.
\end{abstract}

 
\section{Introduction}

  When two gaseous galaxies merge, the gas sinks towards the 
 center of the merger product, creating an intense burst  of new stars,
 and the merger product has the surface brightness
 profile dominated by the $r^{1/4}$ law (e.g., Mihos \& Hernquist 1994).
  Thus, the detection of many smooth, highly-concentrated, blue star-forming
 galaxies would provide strong support for the formation of early-type 
 galaxies via mergers at $z < 1$, as predicted by some 
 hierarchical merger models (e.g., Baugh et al. 1996; Kauffmann et al. 1996).

  Recently, several groups have identified blue 
 galaxies at $z \lesssim 1$ that appear to be  
 early-type galaxies such as E or S0s (Im et al. 2001; Menanteau et al.
 1999; Schade et al. 1999; Abraham et al. 1999; Franceschini et al. 1998). Their 
 surface brightness profiles show a significant $r^{1/4}$ component typical of
 nearby E and S0 galaxies, typically with bulge-to-total light ratios ($B/T$)
 greater than 0.4 (Im et al. 2001). 
  Some of these galaxies are found to have [OII] emission lines,
 indicative of recent star formation (Koo et al. 1996; Schade et al. 1999). 
  Also, others find younger, blue stellar components 
 on top of underlying old, red stellar populations (Menanteau et al. 2000;
 Abraham et al. 1999). 
   These morphological features, as well as their blue colors 
 and the rather high luminosities of at least some of the blue spheroidal
 galaxies, suggest 
 that they may be direct progenitors of
 the present-day typical, massive early-type galaxies
 ($> 10^{11} M_{\sun}$ or $\sigma > 150$ km/sec)
 which are undergoing strong star formation after merging.

  The blue spheroidal galaxies, however, could instead be low-mass,
 starbursting dwarf galaxies such as dwarf ellipticals/spheroids (hereafter,
 dE following Mihalas \& Binney 1981; e.g., Im et al. 1995a, 1995b; 
 Babul \& Ferguson 1996; Driver et al. 1996).
  The strong star formation activity at the center
 might create excess light over the exponential surface brightness
 profile commonly found in dwarf spheroids (Lin \& Faber 1983;
 Gallagher \& Wyse 1994), 
 which could be interpreted as a significant bulge component.
  In fact, some blue spheroidal galaxies resemble 
 distant compact galaxies (CGs; Koo et al. 1994, 1996; Guzm\'an et al. 1996,
 1997, 1998; Phillips et al. 1997).  The previous works on CGs suggest
 that at least some CGs are likely to be progenitors of low-mass spheroids
 ($\lesssim 10^{10}\, M_{\sun}$) today.
  
  As a part of the DEEP Groth Strip Survey (GSS; Vogt et al. 2001a; 
 Koo et al. 1996), we have identified 10 
 BSCs from HST images, and we have measured their emission line widths 
 to obtain mass estimates.
  In this paper, we present evidence that,
 based on their kinematic and structural properties,  
 the majority of BSCs are likely to be low-mass spheroidal galaxies
 undergoing strong starbursts, rather than massive star-forming E/S0s.  
  Throughout the paper, the rest-frame quantities 
 and model predictions are calculated assuming $\Omega_{m}=0.2$,
 $\Lambda=0$ and $h=0.7$ for the cosmological parameters,
  where the Hubble constant is normalized as
 $h=H_{0}/100~$km/sec.

\section{Data}

\subsection{Sample}

  The BSCs are drawn from a set of 28 contiguous HST WFPC2 fields,
 the so called ``Groth Strip'' (Groth et al. 1994; Rhodes et al. 2000).
 A significant portion of the Groth strip was spectroscopically
 observed with the Keck 10-m telescope
 by the Deep Extragalactic Evolutionary Probe
 (DEEP; Koo et al. 1996; Vogt et al. 2001a; Simard et al. 2001; 
 Phillips et al. 2001).

  A combination of HST images and Keck spectra enables us to select BSCs  
 as described below.  We emphasize that our selection criteria for BSCs
 are quite different from those of the CGs in Phillips et al. (1997) and 
 Guz\'man et al. (1997), and that 
 our goal is to select galaxies which morphologically resemble local Es and S0s
 but with blue colors. 

  The first step in our procedure is to fit each HST galaxy image
 using the 2-dimensional bulge+disk 
 surface brightness fitting algorithm GIM2D
 (Simard et al. 1999, 2001; Marleau \& Simard 1998). 
  The main purpose of fitting HST galaxy images with GIM2D is to derive 
 structural and morphological parameters that can be used for 
 quantitative selection of E/S0 galaxies (Im et al. 2001).
  The surface brightness profile of the bulge component is fitted with either  
 a de Vaucouleurs (or $r^{1/4}$) profile or an exponential profile.
  The de Vaucouleurs profile is known to fit the surface brightness profile  
 of ellipticals and bulges of luminous spiral galaxies
 (de Vaucouleurs 1948). Bulges of 
 less luminous, late-type spirals are known to be fitted well with 
 an exponential profile (e.g., 
 Andredakis \& Sanders 1994; de Jong 1996).
  For the component fitted as a bulge,
 we will use the term ``photo-bulge'',  since this component is 
 mathematically defined by the surface photometry, though in some cases
 may not correspond to physically-defined ``bulges'' of nearby galaxies 
 in terms of stellar population and kinematics.
  For the sample selection of BSCs, we will only use a deVaucouleurs profile
 for the photo-bulge component.
  The surface brightness profile of the disk component is fitted
 with an exponential profile, which is known to fit the SB of disks of
 spiral galaxies. In analogy to the photo-bulges, we will use 
 the term ``photo-disk'' for this component.  
  GIM2D measures model-fit structural parameters
 which include the total magnitude; bulge-to-total light ratio ($B/T$); 
 disk scale length
 ($r_{d}$); bulge effective radius ($r_{e}$);
 galaxy half-light radius ($r_{hl}$); position 
 angles and ellipticities for both the bulge and disk components,
 and errors associated with each quantity. Centers of
 both the photo-bulge and photo-disk components are assumed to be the same,
 and fitted along with the other structural parameters. This procedure of
 surface brightness fitting is similar to that of Ratnatunga, Griffiths,
 and Ostrander (1999), and Schade et al. (1995). 
 GIM2D searches for a parameter set which gives the maximum likelihood
 value using Monte-Carlo sampling of the parameter space. 
  Then, GIM2D  Monte-Carlo samples the region around the maximum likelihood, 
 and the median value of each of the parameters is adopted as the 
 model-fit parameter; the errors are estimated at the 68\% confidence level of
 the statistical distribution of the sampled parameter sets
 (Simard et al. 2001). The errors may be underestimated for two
 reasons: (1) PSF modeling errors are not included; (2)
 the sky levels are not fitted along with other parameters, 
 thus the confidence intervals do not include errors on the sky 
 level estimate. Our tests with simulations show, however, that
 these effects do not severely underestimate errors (see  
 Simard et al. 2001).  The assumed luminosity profiles can also affect 
 the output parameters (see section 3.2).
  In general, $V$ and $I$ band images are fitted with GIM2D separately,
 which we call as ``separate fit'', as opposed to ``simultaneous fit''
 where $V$ and $I$ band images are fitted simultaneously to 
 provide better constraints on the model-fit parameters such as
 $V-I$ (see section 3.2). 

  In the next step, early-type galaxy candidates are selected
 based on two quantitative morphological parameters, the bulge-to-total
 light ratio ($B/T$) and the residual parameter ($R$), which are derived 
 from GIM2D using the ``separate'' fits. We use the ``separate'' fits 
 for the sample selection since our early-type galaxy candidates
 come from the early-type galaxy sample of Im et al. (2001), 
 which is based on the ``separate'' fit.
  The quantity $B/T$ measures the 
 ratio of the luminosity in the photo-bulge component
 to  the total luminosity contained in both photo-bulge
 and photo-disk components.
  Thus, the $B/T$ describes 
 how prominent the photo-bulge component is, and typical values are 
 $B/T > 0.5$ and $B/T > 0.3$ for visually classified 
 local Es and S0s, respectively (Im et al. 2001; Kent 1985; Scorza et al.
 1998).
  Note that the $B/T$ cut alone is not very effective at isolating
 E/S0s. Some late-type galaxies  have $B/T > 0.3$ -- 0.4, and 
 about 30\% of objects selected with the $B/T > 0.3$ -- 0.4 cut are 
 indeed non-E/S0 galaxies (Im et al. 2001; Kent 1985).
  For that reason, we use another parameter --
 the residual parameter ($R$) -- 
 to exclude non-E/S0 galaxies with large $B/T$ values.

  The residual parameter $R$ describes how 
 symmetric and prominent the morphological 
 substructures are (such as spiral arms and HII regions).
   Smooth-looking, symmetric galaxies like 
 E/S0s generally have $R \lesssim 0.08$. For a detailed description of 
 the $R$ parameter, see Im et al. (2001) and Schade et al. (1995).
 Im et al. (2001) select ``Quantitatively-Selected E/S0s''
 (hereafter, QS-E/S0s) with the criteria   $B/T > 0.4$
 and $R \lesssim 0.05$ -- $0.08$, depending on the 
 size and apparent magnitude of the object.  Tests on simulated 
 local galaxy images show that these criteria select E/S0s well
 (Im et al. 2001).
  The $R$ cut varies as follows: 
 $R \leq 0.08$ when $I < 21.0$, $R \leq 0.06$ when
 $21.0 < I < 21.5$, and $R \leq 0.05$ when $21.5 < I < 22$. 
  ($I$-band magnitude is defined as the total model magnitude
 derived from the surface brightness fit; 
 note that this total magnitude
 will be used throughout the paper, 
 unless noted otherwise).
  The $R$ cuts are chosen to minimize the contamination 
 of the QS-E/S0 sample by late-type galaxies 
 when the S/N is low ($S/N \lesssim 50$).
  We use the same criteria to choose  QS-E/S0s in this paper, which 
 are identical to QS-E/S0s in Im et al. (2001).
   
\begin{figure}[th]
\psfig{figure=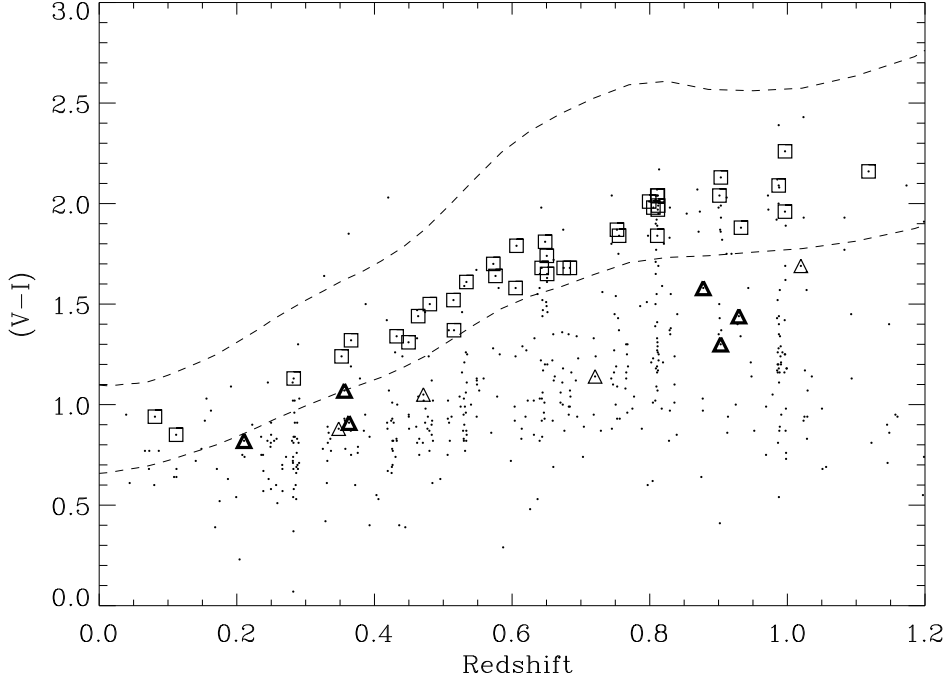,width=13.cm,angle=90}
\figcaption[im.fig1.ps]{
  Spectroscopic redshift vs. $(V-I)$ color for 262 GSS galaxies
 with $16.5 < I < 22$ (dots).
  QS-E/S0s are shown as open squares and triangles.
 Among them, BSCs are marked with thick triangles, and additional
 ``possible'' BSCs are marked with thin triangles (see text).
  The two dashed lines represent the upper and lower limit
 on colors of a passively evolving stellar population formed at $z_{for}=11$
 (for more details, see text). BSCs are chosen as QS-E/S0s which do not 
 meet this color criterion.
\label{fig:zvi}
}
\end{figure}

   As a next step, the spectroscopic redshift-color diagram is used to 
 identify unusually blue QS-E/S0s as BSCs,
 where the color ($V-I$) is also derived from the model total magnitudes.
  Fig. \ref{fig:zvi} shows the $(V-I)$ vs. spectroscopic redshift diagram of 
 a magnitude limited sample ($I < 22$) of 262 galaxies in the Groth strip.
 Note that the identification of spectroscopic redshifts
 is virtually complete to this magnitude limit for the GSS 
 (Phillips et al. 2001).
   Also shown in Fig. \ref{fig:zvi} are lines of plausible color ranges
 for passively evolving, old stellar populations computed with
 the 1996 version of the spectral synthesis model of Bruzual \& Charlot (1993;
 hereafter, GISSEL96) -- 
 the upper dashed line is created by assuming a 0.1 Gyr burst 
 with 250\% solar metallicity,  Salpeter IMF, and  formation 
 redshift $z_{for}=11$.
  The lower line assumes the same model with lower metallicity (40\% solar).
  To allow for magnitude errors, we add/subtract 0.15 magnitude
 in $(V-I)$ color to the upper/lower lines.
  Im et al. (2001) find that  most  QS-E/S0s (marked with squares)   
  form  a red envelope in the $(V-I)$ vs. redshift diagram.
  However, six out of 44 QS-E/S0s  
 (marked with thick triangles)  
 have colors that are bluer than the expected color range of
 the passive evolution model; 
 we call these QS-E/S0s ``good'' BSCs.
 
\begin{figure}[bh]
\psfig{figure=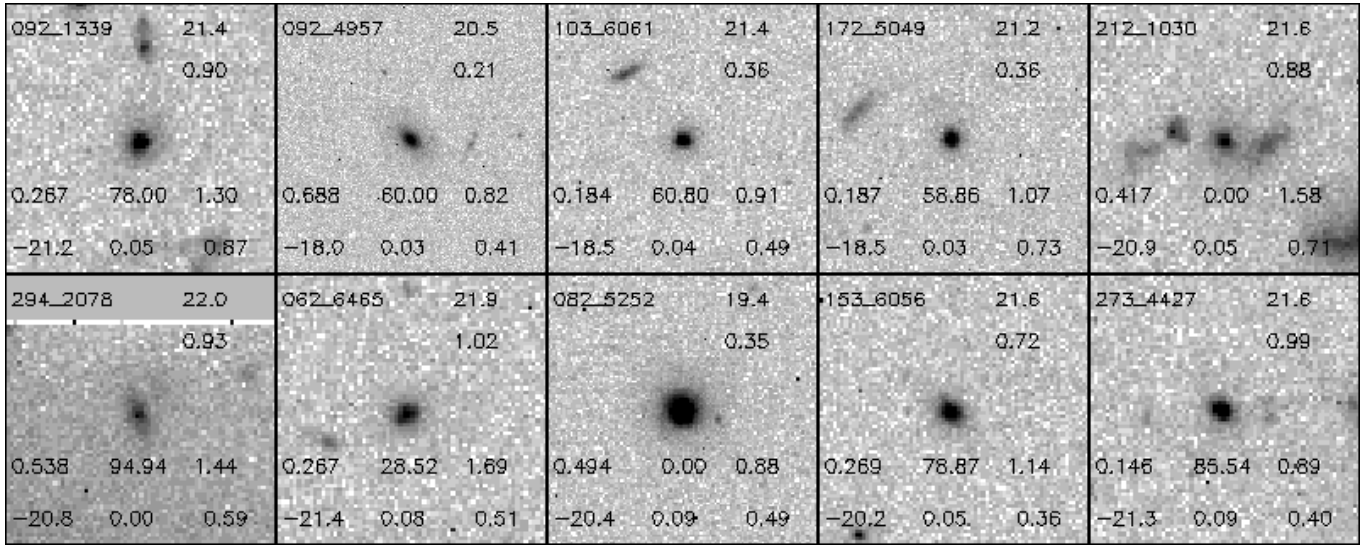,width=18.0cm,angle=0}
\figcaption[im.fig2.ps]{
  Postage stamp images of 10 BSCs ($I$-band). The first 6 images (from
 top left to right) are for ``good'' BSCs, and the remaining images
 are for ``possible'' BSCs.
 Parameters listed in Fig. \ref{fig:besoimage} 
 are, from left to right, and from top to bottom,
 (1) ID\# in our GSS catalog, (2) $I_{tot}$ magnitude,
 (3) redshift, (4) effective radius $r_{e}$(arcsec),
 (5) velocity dispersion $\sigma$,
 (6) observed $(V-I)$, (7) $B$-band absolute magnitude, 
 (8) residual parameter $R$, (9) $B/T$. Boxes are scaled to 
 be 50 kpc on its side.
\label{fig:besoimage}
}
\end{figure}

  In addition to these ``good'' BSCs, 
 we have examined images of galaxies that would have been selected 
 as BSCs if the morphological selection criteria were slightly less
 restrictive ($B/T > 0.35$  or $R \lesssim 0.09$). 
  This visual inspection reveals 4 additional candidates, 
 and we call them ``possible'' BSCs (thin triangles in  Fig. \ref{fig:zvi}).
  Fig. \ref{fig:besoimage} shows $I$-band images of the 6 ``good'' BSCs 
 and 4 ``possible'' BSCs. 
  By constructing an azimuthally-averaged surface-brightness (SB) 
 profile using ellipsoidal apertures and plotting it along the 
 major axis,  we confirm that our BSCs show surface brightness profiles 
 which differ significantly from
 a simple exponential law, but that are well fitted by the addition of 
 a significant photo-bulge component.
  This and other important points regarding the colors and structures of
 the underlying stellar populations are discussed in more detail
 in Section 3.2.

\begin{figure}[th]
\psfig{figure=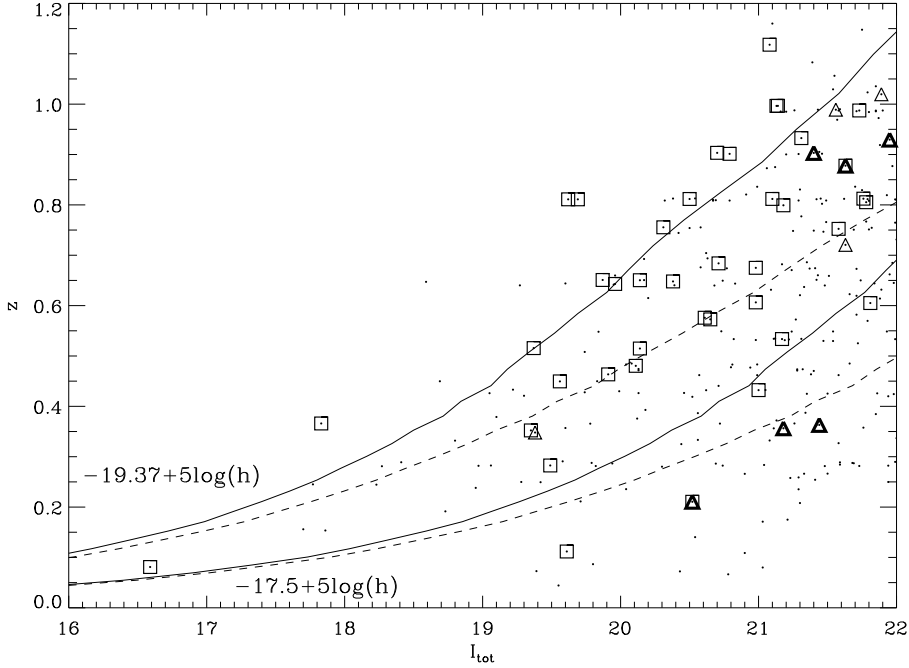,width=12.5cm,angle=90}
\figcaption[im.fig3.ps]{
 Redshift vs. $I$ magnitude of GSS galaxies. 
 QS-E/S0s are marked with squares, ``good'' BSCs are marked with
 thick triangles, and ``possible'' BSCs are marked with thin triangles.
 The remaining galaxies are plotted as dots. 
 Lines indicate trajectories of galaxies with two different 
 absolute magnitudes at $z=0$: 
 $M_{B}(z=0) = -19.37 + 5$ log$(h)$ ($\sim L^{*}$; Marzke et al. 1998) and
 $M_{B}(z=0) = -17.5 + 5$ log$(h)$. 
  The solid lines assume passive luminosity evolution 
 equal to $-1.7 \times z$ in $B$-magnitudes (Im et al.
 2001; Gebhardt et al. 2001; Koo et al. 2001), while dashed lines do 
 not take the evolutionary correction into account.
  For the $K$-correction
 of the non-evolving models, we use the 13 Gyr-old SED of 
 a model stellar population with 0.1 Gyr burst SFR, $z_{for}=11$, 
 and the Salpeter IMF (GISSEL96).
\label{fig:zi}
}
\end{figure}

  Fig. \ref{fig:zi} shows the spectroscopic redshifts vs. $I$ for  ``good''
 BSCs (thick triangles) and ``possible'' BSCs (thin triangles),
 compared with red QS-E/S0s (squares) 
 and all other types of galaxies (dots) 
 in the GSS.
  Below $z=0.5$, there tends to be a dichotomy in the sense that 
 red QS-E/S0s are intrinsically bright while BSCs are faint. 
  Their faint apparent magnitudes 
 suggest that low-$z$ BSCs are unlikely to evolve into massive E/S0s today.  
  In contrast, above  $z = 0.5$, BSCs approach $L^{*}$ in luminosity. 
  However, the apparent brightness of $z > 0.5$ BSCs does not
 necessarily mean that they will evolve into the present-day 
 $L^{*}$ E/S0s, 
 as the luminosity dimming for these galaxies may be more significant
 than that of the 
 old stellar populations which are assumed in lines of Fig.\ref{fig:zi} 
 -- thus, these $L \sim L^{*}$ high-$z$ BSCs may fade significantly 
 into low-luminosity systems today. 
   
   To improve constraints on the local counterparts of BSCs, 
 we need to study quantities 
 which are independent of the luminosity evolution and are more direct 
 tracers of the mass, e.g.,
 the internal velocity dispersion $\sigma$.
 The next section describes our measurements
 of the internal velocity dispersion, $\sigma$, for these BSCs.

\subsection{Spectroscopic observations and line width measurements}

  Spectra of BSCs were obtained as a part of the DEEP survey 
 during the 1996-1999 Keck observing runs (Phillips et al. 2001) 
 using the Low Resolution Imaging Spectrograph (LRIS; Oke et al. 1995).
 A 900 lines/mm grating was used to cover the blue part of 
 the spectrum ($\sim 5000$ -- $6500$ \AA) and a 600 lines/mm grating was
 used for the red part ($\sim 6500$ -- $8500$ \AA); typical exposures
 were $\sim$ 1 hour with each grating.
  The spectral resolution 
 (FWHM$\sim$ 2 \AA~ and 4 \AA~ for the blue and
 red parts, respectively, or $\Delta \sigma \sim 80$ km sec$^{-1}$
 with $\Delta \sigma$ defined as $\sigma$ of the instrumental 
 resolution)
 allows us to measure $\sigma$ as low as $\sim$ 30 km\, sec$^{-1}$. 
  For one BSC (092\_1339), we also obtained a 15-min high resolution
 spectrum  
 ($\Delta \sigma \sim 17$ km sec$^{-1}$) using the Echellette Spectroscopic
 Imager (ESI; Sheinis, et al. 2000) during May, 2000, at Keck. 

  To measure the line widths, we fit Gaussian profiles to 
 strong S/N emission lines  
 following the method described in Phillips et al. (1997) and
 Guzm\'an et al. (1997).
  For the [OII] doublet at 3727 \AA, we used a double Gaussian 
 profile with a fixed $\Delta \lambda$ between both peaks. 
  Each velocity width was corrected for the instrumental resolution. 
  Finally, we take the variance-weighted mean value of the velocity widths
 measured from different lines as our adopted $\sigma$.

  Estimating true errors of velocity widths is not a trivial task, as 
  uncertainties in velocity widths include various factors beyond
 the formal measurement error.
  This is particularly true in estimating the instrumental
 correction, which varies with galaxy extent, seeing disk size, and
 misalignment of the target with the slit.  In addition, the non-uniform
distribution of the emission and non-Gaussian velocity profiles
contribute to the intrinsic error. Simulations with realistic
 parameters indicate a ``robust'' velocity error of about one-third the
 instrumental line width, a value which is supported by repeat observations
 of the same objects, and by line width measurements using different algorithms
 (Phillips et al. 2001; Guzm\'an et al. 1997).
 For LRIS, this corresponds 
 to $\sim 30$ km sec$^{-1}$ depending on the line and redshift 
 of the object.  Hence, we adopt 1-$\sigma$ error, 
 $\delta \sigma = 30$ km sec$^{-1}$, as
 a good (conservative) error estimate (see Phillips et al. 2001 for
a full discussion). 
  For the ESI data, we adopt
 $\delta \sigma = 6$ km sec$^{-1}$, which is simply one-third of the 
 ESI resolution. The $\sigma$ values can be found in Table 1.

\begin{figure}
\psfig{figure=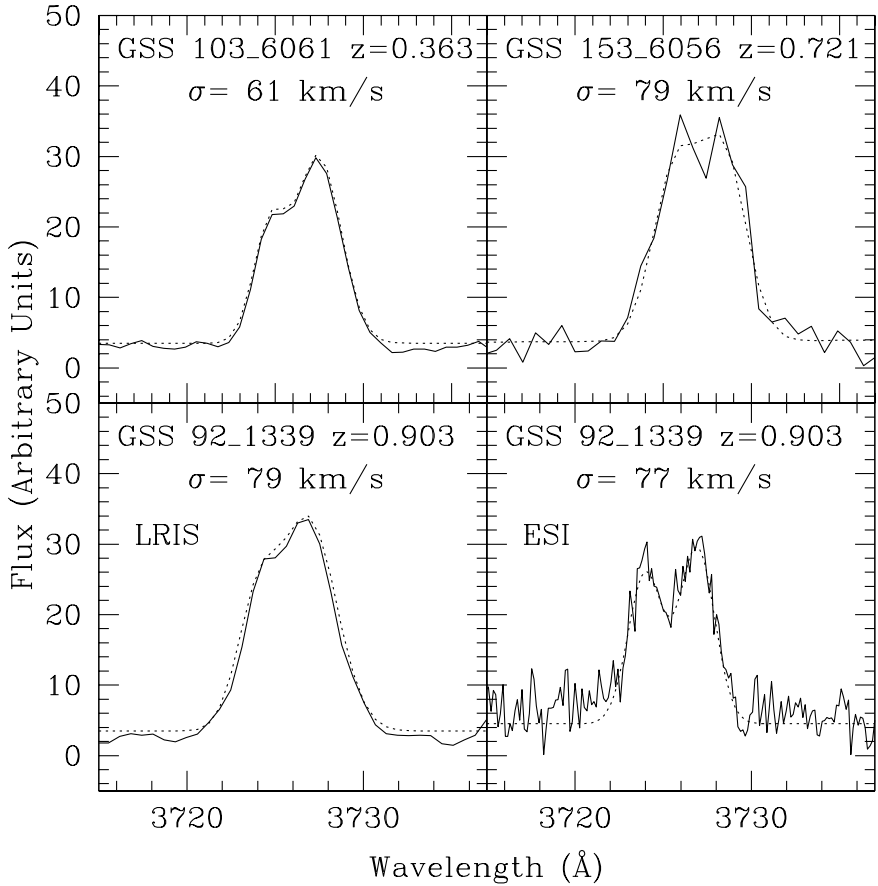,width=12.3cm,angle=0}
\figcaption[im.fig4.ps]{
 The OII [3727] doublet for representative BSCs. Most spectra here 
 come from LRIS data. For 092\_1339, 
 we also show  ESI data. The comparison between the LRIS and  
 ESI data supports the
 accuracy of the line width measurements from lower-resolution LRIS data.
\label{fig:spectra}
}
\end{figure}

  In Fig. \ref{fig:spectra}, we show spectra centered on the [OII] emission
 line for 3 representative BSCs.
 Interestingly, an inspection of the 2-D spectra of two ``good'' BSCs
 (092\_4957 and 294\_2078) 
 show tilted emission lines, indicating that they are rotating objects
 (e.g, disks). 
  One ``good'' BSC (212\_1030) has a weak [OII] emission line,  
 but it is not clear whether the emission line comes from the object
 itself or from its blue neighbors, at least one of which 
 is at the same redshift.
  Finally, one ``possible'' BSC (082\_5252) has a redshift taken from 
 the literature (Lilly et al. 1995), so    
 we do not have a measured $\sigma$ for this object.

\section{Results}

  In this section, we present our two main findings about 
 BSCs. In section 3.1, we will show that the dynamical masses of BSCs
 are similar to those of nearby low-mass spheroidal galaxies, assuming
 that the dynamical mass estimated from the emission line widths is not 
 severely underestimated (less than a factor of $\sim 2$). 
  Then, in section 3.2, we show that 5 -- 6 out of 10 BSCs seem to have
 an underlying  
 old stellar population, and that their blue overall colors originate from 
 localized regions containing young stars.  In the same section, 
 we also discuss a plausible shape of the surface brightness profile
 as a means to explore the structure of the underlying stellar population.

\subsection{Mass of BSCs}

  In Table 1, we present the basic data for BSCs. Note that structural
 parameter values in Table 1 come from the ``separate'' fit described 
 in Section 2. 
   We estimate the dynamical mass 
 using the virial theorem for spherical $r^{1/4}$ law systems 
  (Poveda 1958; Young 1976; Illingworth 1976):

  \[ M_{dyn} \simeq 3 \sigma_{3D}^{2} r_{1/2} / G, \]

\noindent 
  where $\sigma_{3D}$ is the central velocity dispersion of the system
 in real 3-D space, $r_{1/2}$ is the projected half-mass radius, 
 and $G$ is the gravitational constant. 
  We assume that stellar mass traces light, i.e.,
  $r_{1/2} \simeq r_{hl,cir}$, where $r_{hl,cir}$ is the projected circular 
 aperture half-light radius of the stellar system in Table 1
 (roughly equal to the major-axis half light radius,
 if the ellipticity, $e$, is small).
   Note that we use $r_{hl,cir}$, instead of $r_{hl,maj}$ which 
 is defined along the major axis and given as a default output from
 GIM2D, in order to facilitate the comparison
 with local samples for which $r_{hl,maj}$ is not readily 
 available.   
  The line of sight velocity dispersion $\sigma$ would be 
 $\sigma_{3D}/ \sqrt{3}$ in an isotropic velocity field. Therefore, the 
 total dynamical mass of the system is roughly,

\begin{equation}
 M_{dyn} \simeq 9 \,\sigma^{2} \,r_{hl,cir}/G.       
\end{equation}

\noindent

  This formula is similar to that adopted by Phillips et al.
 (1997). 

\begin{figure}
\psfig{figure=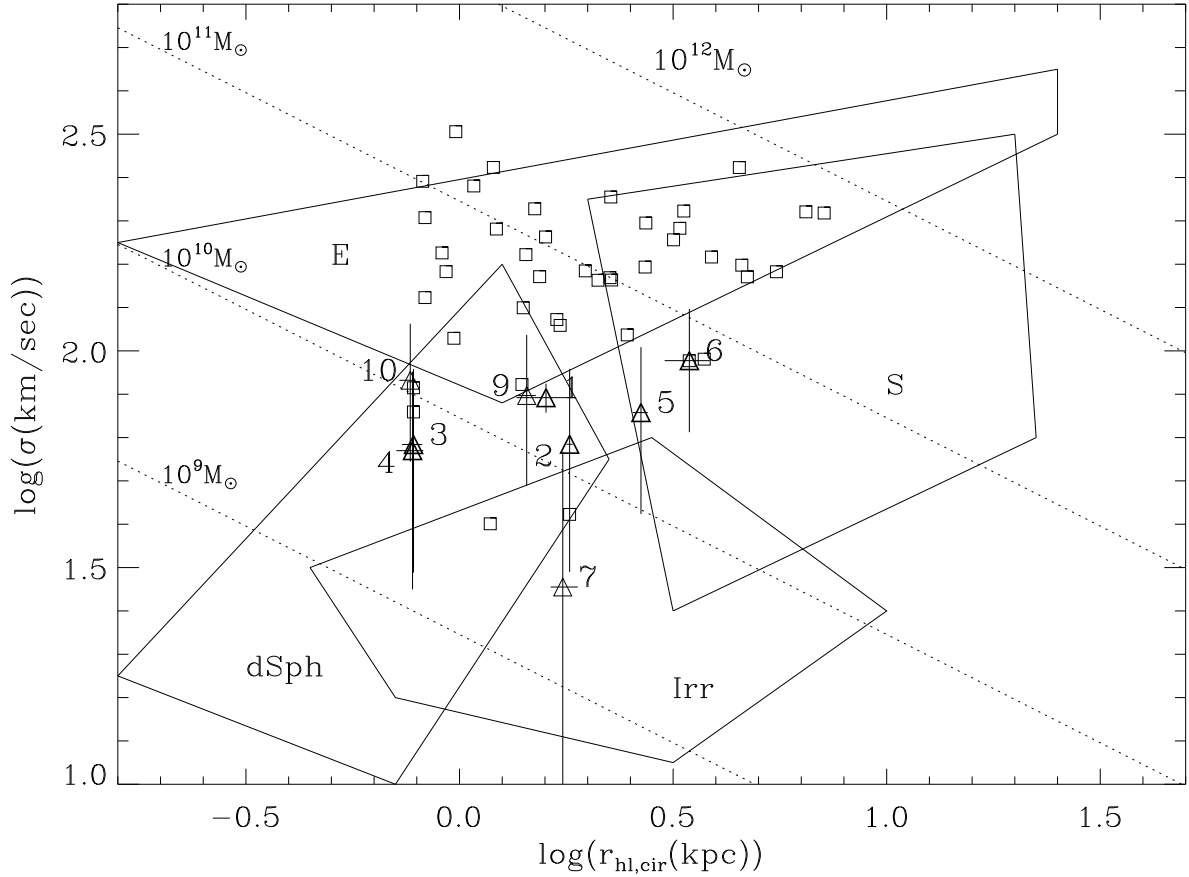,width=17.3cm,angle=90}
\figcaption[im.fig5.ps]{
 $\sigma$ vs. $R_{hl,cir}$ for BSCs.
 The spaces occupied by typical local Hubble types are taken 
 from Phillips et al. (1997). BSCs (triangles) are generally low-mass objects 
 that lie in or near the space
 occupied by  dEs).
  Red, QS-E/S0 galaxies with spectroscopic redshifts in Im et al. (2001)
 are plotted with squares showing a region where these QS-E/S0s
 with $I < 22$ are expected to be found.  
  The number plotted next to each point is the object number that 
 appears in the first column of Table 1. The dotted lines show loci 
 of constant mass from Eq. (1).  Object \#8 is not plotted as we have no $\sigma$ measured for it.
\label{fig:resig}
}
\end{figure}

   Fig. \ref{fig:resig},  
   shows measured $\sigma$ vs. $r_{hl,cir}$ for all BSCs (triangles).
  Also plotted in Fig. \ref{fig:resig} are regions
 occupied by different galaxy types at 
 $z=0$, taken from Phillips et al. (1997) and
 Guzm\'an et al. (1998).  Dashed lines are drawn to indicate  constant
 mass. In the same plot, squares indicate $r_{hl,cir}$ vs. $\sigma$ of
 distant, red, QS-E/S0s ($I < 22$) with spectroscopic redshifts  
 in Im et al. (2001). 
  For the Im et al. (2001) sample, we use measured $\sigma_{abs}$
 when they are available (Gebhardt et al. 2001), and $\sigma_{FP}$ 
 (see Section 4.1) otherwise.
   Fig. \ref{fig:resig} shows that no BSCs resemble
 present-day, massive, early-type galaxies or distant red, QS-E/S0s.
  The majority of BSCs (5 out of 9) lie within or on the border of 
 the polygon for local dEs, suggesting a possible link.
  BSCs have $\sigma \lesssim 80$ km/sec and $M \lesssim 
  10^{10} M_{\sun}$. 
  Typical, massive, luminous  E/S0s ($L \gtrsim L^{*}$) have 
 $\sigma \gtrsim$ 150 km/sec and $M \gtrsim  10^{11} M_{\sun}$.
  Thus, BSCs are {\it not} likely to evolve into massive E/S0s today 
 under the assumption that the measured $\sigma$  
 reflect the dynamical galaxy mass. Studies exist which 
 support this assumption (e.g., Telles \& Terlevich 1993; 
 Kobulnicky \& Gebhardt 2000 and references therein), but 
 future studies with more extended samples may be needed 
 to establish the link between $\sigma$ measured from emission lines
 (hereafter, $\sigma_{em}$) and dynamical mass of galaxies --  
 these points are discussed in section 4.1, where we argue  
 that our conclusion here is not likely to change
 even if our $\sigma_{em}$ values somewhat underestimate $M_{dyn}$. 

  Note that CGs studied by Phillips et al. (1997)
 and Guzm\'an et al. (1996) also lie in a similar region where
 BSCs are (see Fig.9 of Phillips et al. 1997).
  From the $B/T$ and $R$ values which we derived
 separately for CGs, 
  we find that roughly 10\% of CGs at $I < 22$ in the Phillips et al. (1997)
 and Guzm\'an et al. (1996) sample can be classified as BSCs.
  The number of BSCs found in the CG sample per WFPC2 frame (0.25 per frame)
 matches well with the number of BSCs found here.
  Conversely, we can apply the selection criteria of CGs
 in Phillips et al. (1997) to our sample,
 and we find that 8 out of 10 BSCs are CGs.
  Thus, we conclude that most BSCs belong to a subset of CGs that are  
 smooth and photo-bulge-dominated.

   A more detailed description of each BSC in Fig.\ref{fig:resig} is  
 presented below.
   One BSC (062\_6465; \#7) lies in the region of Irr galaxies, but not too 
 far from the locus of dEs. 
   Two BSCs (294\_2078, 212\_1030; \#6 and \#5 respectively)
 are found in the spiral galaxy 
 polygon.  As noted above, the spectrum of one of them (294\_2078)
 shows tilted emission lines
 indicative of a rotating system. 
  Furthermore, the measured $B/T$ value has a relatively large error
 (e.g., $B/T=0.59^{+0.07}_{-0.14}$ from Table 1),
 and this galaxy is a highly-inclined 
 system. Thus, BSC 294\_2078 is probably an early-type spiral with
 a significant disk component or a later-type galaxy.
  BSC 212\_1030 is an absorption-line object with a possible weak [OII]
 emission line. The blue companion on the left in Fig. 2 has a redshift
 virtually identical to  212\_1030 ($z=0.88$),
 and is possibly physically associated.
 The projected separation between both objects is
 $\sim$ 1.2 arcsec (6 $h^{-1}$ kpc). 
 The blue galaxies on the right are foreground objects with redshifts
 $z \simeq 0.24$, much
 lower than the redshift of 212\_1030, and so are  not 
 physically  associated.
 Unfortunately, the spectrum of 212\_1030 is too noisy to 
 measure the absorption-line velocity width. 
  Its nuclear color measured with an aperture size of 0.5 arcsec is 
 $V-I = 1.9$, much redder than the total color $V-I = 1.58$ from the 
 model fit. We also note that model-fit colors of this object
 using different procedures are $V-I \simeq 1.7$ -- $1.8$. 
 Therefore, we suspect that the model-fit color of 212\_1030 is 
 somehow contaminated by its blue neighbors, 
 and that  this object is not a genuine BSC, but rather is more likely 
 to be a normal, red, early-type galaxy. 

   Another BSC (092\_4957; \#2), which lies close to the border of
 dEs and spirals,  has a rotation curve with
 an inclination-corrected rotational velocity, $V_{max} \simeq 130$ km/sec
 (Vogt 2000, private communication).  The velocity width of $\sigma \simeq 60$
 km sec$^{-1}$,  is in agreement with the $V_{max}$ value considering  
 the conversion factor of $\frac{1.81}{sin(inc)}$ in 
 $V_{max} \simeq \frac{1.81}{sin(inc)} \times \sigma_{[OII]}$,  
 which typically amounts to $\sim 2$ (Kobulnicky \& Gebhardt 2000).
  The disk component of this object is found to  
 lie on the local TF relation, a strong indication that 
 this object may be a relatively low-mass spiral galaxy
 (Vogt et al. 2001b).

\subsection{Underlying stellar population}

\begin{figure}[hb]
\psfig{figure=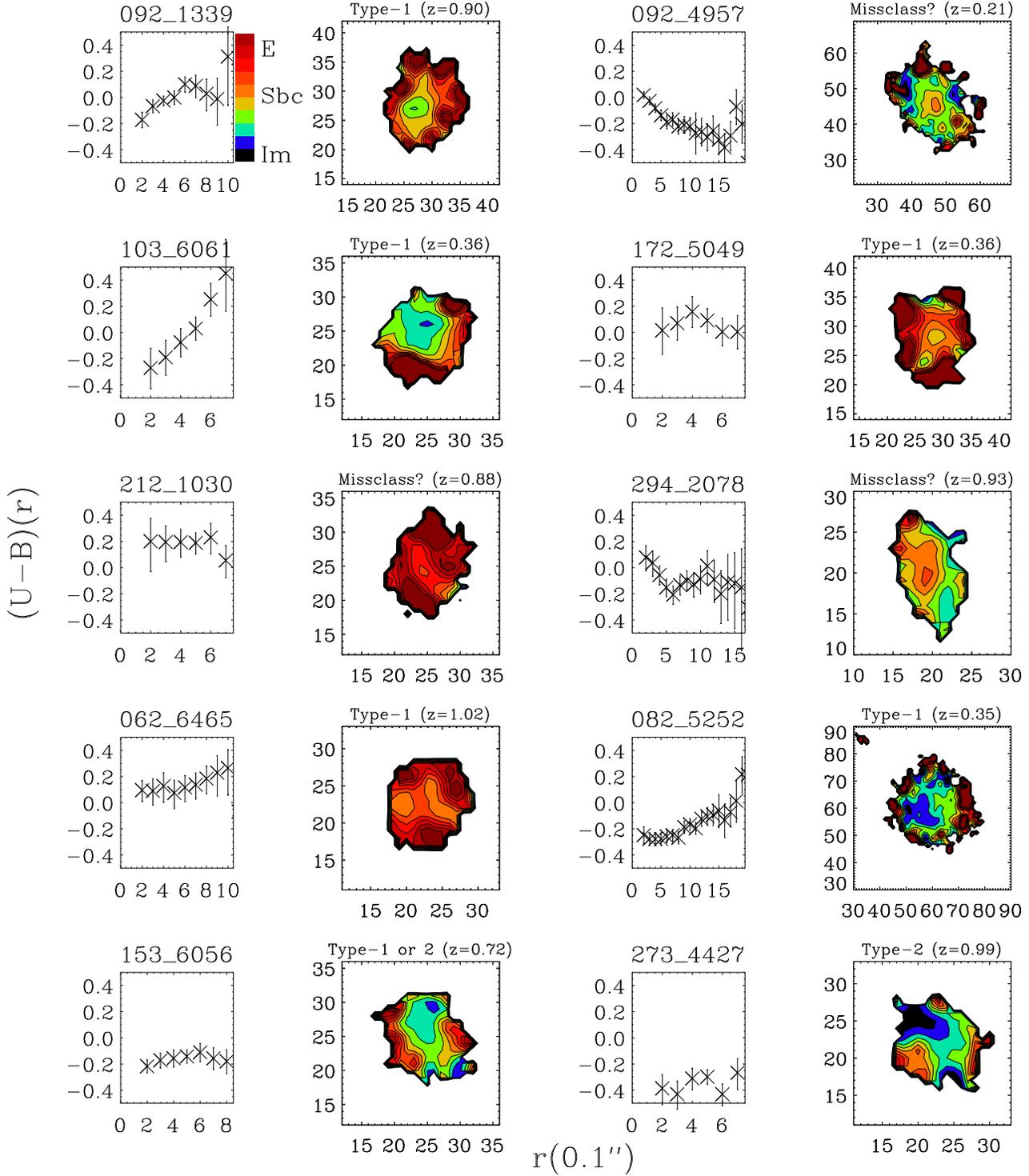,width=17.cm,angle=0}
\figcaption[im.fig6.ps]{
  Azimuthally averaged rest-frame $U-B$ color profiles plotted 
 along their major axes, and 
 2-dimensional $U-B$ color maps of BSCs. 
  For the 1-D color profile, the x-axis shows  
 the distance from the center of the object, $r$, in 0.1 arcsec.
  For the 2-D color map, the x and y axes are also in units of 
 pixels. 
  Type-1 BSCs have a central region significantly 
 bluer than the outer region, which in many cases is almost as red as 
 typical field E/S0 galaxies. Note that the expected $U-B$ colors of
 passively evolving stellar populations after 0.1 Gyr bursts range 
 from 0.1 to 0.5 depending on the age and metallicity at the
 redshift of interest (see text). 
\label{fig:grad}
}
\end{figure}

  The accretion of gas-rich satellite galaxies or gas infall toward
 the centers of 
 E/S0 galaxies may provide fuel which will trigger intense star
 formation. Young stars in the satellite galaxies can also fall into
 the centers.
 The existence of relatively-young stellar populations among local 
 E/S0 galaxies suggests that such a sprinkling of young stars over an old, 
 underlying, stellar population has occurred
 (e.g., Charlot \& Silk 1994; Trager et al. 2000).
  If BSCs are old galaxies sprinkled with young stars near
 their center, we should find that the outskirts 
 are red, while the inner parts are blue.
   Fig. \ref{fig:grad} shows the rest-frame $U-B$ color gradient
 of BSCs along their major axis out to $\sim 3~r_{hl,maj}$, as well as 
  a 2-dimensional color map.
   To construct the 2-D color gradient plot, 
 we used the IRAF \footnote{
 IRAF is distributed by the National Optical Astronomy Observatories, under 
 contract to A.U.R.A. Inc.} 
 task, ``ellipse''  with a fixed object
 center, an ellipticity and a  position angle
 which all correspond to the values derived
 for the photo-bulge component of the ``separate'' 2-D fit.  Background
 sky levels were estimated using a region which is at least 
 5 pixels away from the object pixels \footnote{Object pixels are 
 defined as a contiguous region of 
 pixels that have values $> 1.5 \sigma_{bkg}$ after convolving 
 the image using a 3$\times$3 Gaussian kernel with a FWHM of
 1.5 pixels; see Simard et al. 2001 for more details.}.
  In the process, observed $V-I$ values are converted to
 the rest-frame $U-B$ using 
 the empirical $K$-correction which is presented in Gebhardt et al. (2001).

  To construct the 2-D color map, we used the following procedure:
  First, both the $V$- and $I$-band images of each object are background
 subtracted, and smoothed with Gaussian filter with $\sigma=1$. 
 This smoothing of the image with the Gaussian filter is necessary to 
 enhance the global color gradients.
 In the next step, each pixel value is converted into magnitude units. 
 Then, the $I$-band image is subtracted from the $V$-band image to produce 
 a color image.
  Finally,  the $V-I$ values are converted to the rest-frame $U-B$ color.
  The expected $U-B$ values for a passively evolving stellar population
 after a 0.1 Gyr burst are $U-B = 0.1 \sim 0.2$ at the age of 1 Gyr 
 and $U-B = 0.2 \sim 0.5$ at the age of 5 Gyr, for the assumed metallicities of
 40\% and 100\% solar values, respectively (GISSEL96).


  Fig. \ref{fig:grad} suggests that there are three categories of 
 BSCs, as described below.

 Type-1:  Five, possibly six, out of the 10 BSCs
 show a tendency
 to have blue inner regions and red outer regions, suggesting
 that these BSCs  have underlying old stellar populations 
 ($092\_1339, 103\_6061, 172\_5049, 062\_6465, 082\_5252$, and 
 possibly $153\_6056$).
  Red outer regions  are    
 as red as the expected colors of passively-evolving stellar populations 
 ($U-B \simeq 0.1 \sim 0.5$), 
 and redder than colors of local Sbc's ($U-B \simeq 0.0$).
  Thus, these BSCs might be old galaxies with star formation occurring
 more actively near their centers; similar objects have been discussed  
 in Guzm\'an et al. (1998), Abraham et al. (1999) and Menanteau et al. (2000). 
  Note that Jansen et al. (2000) find that 50\% of the local, low-luminosity
 galaxies have the bluest colors at their centers -- Type-1 BSCs may 
 be comparable to such local low-luminosity galaxies.

 Type-2: One, or possibly two BSCs ($273\_4427$ and possibly
 153\_6056) do not have a significantly 
 redder outer region, and their overall colors are bluer than Sbc's.  

 Misclassifications: Two BSCs have red cores and blue outer regions
 ($092\_4957$ and $294\_2078$).  Both have tilted emission
 lines, indicative of rotation. Their rotation signature, as well as 
 the overall structural and color gradient appearance suggest that
 they are more likely to be late-type galaxies rather than genuine
 BSCs. The remaining  BSC ($212\_1030$) seems to be an early-type
 galaxy with a red color similar to that of local early-type galaxies. 
 
  Further evidence for the existence of red outer regions for Type-1 BSCs  
 comes from the 2-dimensional bulge+disk SB profile fits.
  These provide the structural parameters and 
 colors of photo-bulge and photo-disk components, which can 
 be compared with the results obtained from the color gradient
 analysis.  Since we are interested in obtaining well-constrained colors 
 and structural parameters for each component,
 we use the ``simultaneous'' fitting method, i.e., where 
 we fit the $I$-band and $V$-band
 images of each BSC simultaneously, 
 allowing only the total flux, $B/T$, and the background correction 
 to be different in each band (Simard et al. 2001; Koo et al. 2001). 

  To test the robustness of our results 
 to the assumptions for the adopted  SB profiles, 
 we tried two kinds of SB models -- the ``$r^{1/4}$ + disk model'' 
 which uses a $r^{1/4}$ law profile for the photo-bulge
 and an exponential profile for the photo-disk,
 and the ``double exponential model'' which adopts exponential
 profiles for both photo-bulge and photo-disk components.
  Note that the derived structural parameter
 values from these ``simultaneous'' fits (Tables 2 and 3)
 can be slightly different from
 the values derived from the ``separate'' fits described in section 2,
 which is used  for the sample selection (Table 1). 

\begin{figure}[hb]
\psfig{figure=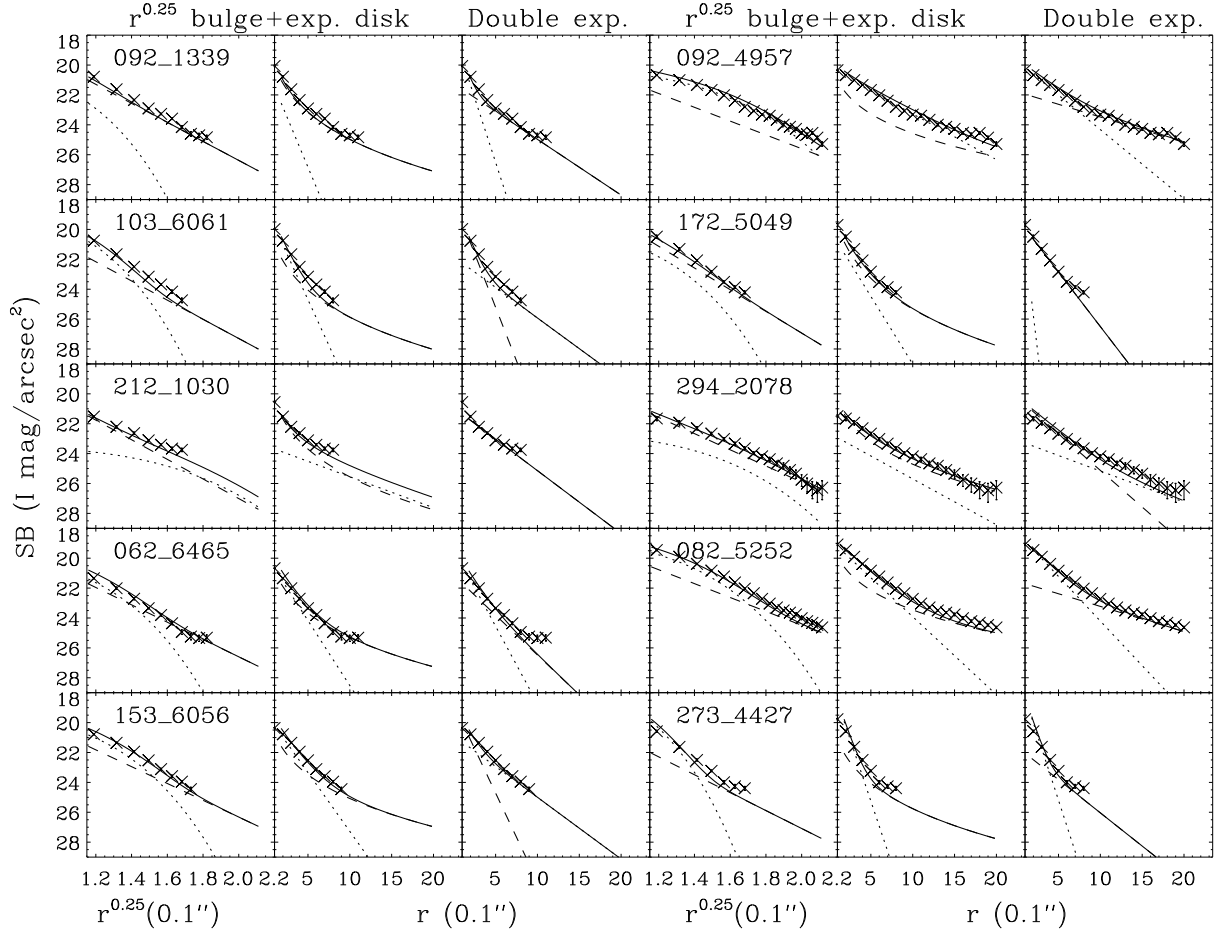,width=17.cm,angle=90}
\figcaption[im.fig7.ps]{
  Azimuthally averaged surface brightness profile of BSCs plotted 
 along their major axis,
 as a function of distance from the center of the object.
  For each object, there are three panels, two for comparison with
 the $r^{1/4}$ bulge+ exponential disk profile,
 with distances indicated in units of
 $r^{0.25}$(0.1 arcsec) and $r$(0.1 arcsec), 
 and a third for comparison with the double exponential profile, 
 with distances given for $r$. 
  A straight line is expected in 
 the $r^{0.25}$ vs. SB plot for the $r^{0.25}$ law profile, while 
 the same is true in the $r$ vs. SB plot for the exponential profile. 
  The ``X'' marks show the measured SB. Models are drawn with the solid line
 (bulge+disk),
 the dashed line (bulge component), and dotted line (disk component), 
 using outputs from the 2-dimensional GIM2D fit.
  Slight discrepancy between these data  and the model exists; 
 this is due to that (1) the aperture data are not deconvolved;
 (2) the aperture photometry is done with a fixed ellipticity
 while the bulge and disk components from the GIM2D fit have different
 ellipticities; and (3) faint neighbors affects the aperture photometry at the 
 very outer regions. 
\label{fig:beso_sb}
}
\end{figure}

  Fig. \ref{fig:beso_sb} shows the azimuthally averaged 
 major axis surface brightness profile  
 of BSCs (points) obtained from the 
 aperture photometry, together with the best-fit model profile from GIM2D  
 (solid line) which is composed of a 
 photo-disk (dotted line) and a photo-bulge (dashed line).
  Fig. \ref{fig:beso_sb} shows that the surface brightness profiles
 {\it cannot} be fitted well with a single exponential profile -- 
  a significant second component is needed to fit their SB profile. 
  In Tables 2 and 3, we list the GIM2D parameters from the ``simultaneous''  
 $r^{1/4}$+disk and double exponential model fits, respectively.
  The important quantity in Tables 2 and 3 is the color of
 the more extended component, i.e., a component
 whose half-light radius is the larger of the two 
 ($r_{hl,maj}=r_{e,maj}$ for a $r^{1/4}$ component,
 and $r_{hl,maj}=1.68 \, r_{d,maj}$  for an exponential component).
   Regardless of the assumed model SB profile,
  the extended components (typically, photo-bulge) of three 
 Type-1 ``good'' BSCs indeed 
 have a color redder than typical Sbcs ($U-B \simeq 0$) or  
 as red as the expected colors of passively-evolving E/S0s
 ($U-B \gtrsim 0.1$), providing an independent confirmation
 for the existence of
 a red underlying stellar population in these BSCs. 
  For the three remaining ``possible'' BSCs, 
 the 2-D profile fits are less conclusive.
  Some redder components are not more extended.
  Nevertheless, even for such cases, the redder and bluer components
 have comparable $r_{hl,maj}$.   We will 
 use the term ``underlying red stellar component'' 
 or ``underlying red component'' to describe 
 the extended red component which seems to exist in 
  Type-1 BSCs.


   In principle, the shape of the surface brightness profile of 
 the underlying red stellar population can tell us about the likely local 
 counterparts of these BSCs. If they are like massive E/S0s, 
 the underlying stellar population will be distributed following 
 the $r^{1/4}$ law profile. On the other hand, the underlying red 
 component will show an exponential SB profile, if BSCs are more likely 
 progenitors of less massive spheroids. 
    Unfortunately,  we find that the difference in reduced $\chi^{2}$ values
 from both fitting models is small ($\Delta \chi^{2} \lesssim 0.01$),
 meaning that the $r^{1/4}$+disk model and the double exponential model are
 almost equally good. 
   Thus, the current data do not constrain 
 the likely local counterpart of the BSCs in terms of
 the surface brightness profile.
 Any underlying old stellar population 
 will stand out more clearly in the near-infrared (NIR; e.g., 
 Hammer et al. 2001); thus an analysis of high-resolution NIR images
 should offer more insights toward the characteristic SB profile
 and the mass content of the red component.

\section{Discussion}

  This section provides detailed discussions on our findings.
  In section 4.1, we discuss the robustness of the use of 
 emission line widths for mass estimates.
  Section 4.2 shows detailed comparison of our BSCs vs. other 
 BSCs in the literature, and section 4.3 discusses BSCs in 
 the Hubble Deep Field.
  A rough estimate of the number density of BSCs is given in section  
 4.4, and section 4.5 discusses BSCs and their interaction with close
 neighbors.

\subsection{Mass of the underlying stellar population}

  In the previous section, we found that 5 or 6 Type-1 BSCs 
 in our sample appear to have an underlying, extended, red stellar
 component ($U-B > 0.0$)
 and a bluer core from where most of the flux in emission lines may originate.
  When the star formation is 
 very localized, the line width from emission lines can be 
 much narrower than a more global velocity dispersion. 
  It is thus reasonable to ask how well the velocity dispersion measured 
 from the emission lines really represents 
 the velocity dispersion of the entire galaxy, especially for
 Type-1 BSCs.
  
  One possible way to resolve this issue is to directly compare 
  $\sigma_{em}$ with the observed absorption-line velocity dispersions
 ($\sigma_{abs}$) of underlying  stellar populations. 
  Kobulnicky \& Gebhardt (2000) measure both $\sigma_{abs}$ and 
 $\sigma_{em}$ for a sample of local galaxies of various
 morphological types, and find that 
 $\sigma_{abs}$ and $\sigma_{em}$ correlate well. However,
 their sample does not contain E/S0 galaxies, and thus may not be directly 
 applicable to BSCs. 
  For a sample of nearby elliptical galaxies, Phillips et al. (1986) 
 measured $\sigma_{em}$(core) from [NII]. 
  We identify 21 ellipticals in Phillips et al. (1986) which overlap with
 the elliptical galaxy sample of Faber et al. (1989) for which   
 $\sigma_{abs}$(core) is available. Here, we use the notation
 ``(core)'', since both $\sigma_{em}$ and
 $\sigma_{abs}$ are measured within $r \lesssim 0.1 \, r_{hl}$. 
  The comparison between $\sigma_{abs}$(core) and $\sigma_{em}$(core)  
 reveals that of these 21 massive ellipticals, all have 
 $\sigma_{abs}$(core) $> 130$ km sec$^{-1}$ and none 
 have $\sigma_{em}$(core)$ < 100$ km sec$^{-1}$.
  Some ellipticals in Phillips et al. (1986) do have   
 $\sigma_{em}$(core) $<$ 80 km sec$^{-1}$, but these galaxies are found to be
 rather less luminous systems ($M_{B} \lesssim -19.5$) 
 and have [NII]$\lambda$6584/$H_{\alpha}$ ratios $\lesssim 0.2$,
 resembling HII galaxies.  Some BSCs have $\sigma_{em}$ and $M_{B}$ values
 very similar to the local HII-like ellipticals.
  A weakness in this comparison is that $\sigma$ are measured only near
 the core in Phillips et al. (1986), while $\sigma_{em}$ of our BSCs are
 measured over a more extended
 region ($\sim r_{hl}$)\footnote{When $\sigma_{em}$ is measured 
 over more extended region as for our BSCs, they are expected to be better
 tracers of the global kinematics.}.
  
  For distant early-type galaxies, 
  Koo et al. (2001) discuss $\sigma_{em}$ vs $\sigma_{abs}$ of luminous bulges
 at $z \sim 0.8$. They find 5 luminous bulges for which $\sigma_{em}$ is 
 measurable, and find a reasonable agreement between $\sigma_{em}$
 and $\sigma_{abs}$. None of these five objects have $\sigma_{em} 
 < 100$ km sec$^{-1}$.
  We find a similar result for the E/S0 galaxy sample of Im et al. (2001), 
 which contains at least 5 objects with both $\sigma_{abs}$ and $\sigma_{em}$.
  
  We would like to emphasize that none of the 
 luminous bulges or E/S0s in either the nearby or distant samples 
 have $\sigma_{em} <$ 100 km sec$^{-1}$.  
  The lack of massive local ellipticals 
 with $\sigma_{em} < 80$ km sec$^{-1}$ suggests that 
 Type-1 or Type-2 BSCs are more likely to be low-mass ellipticals with 
 intense star formation rather than massive E/S0s.

\begin{figure}[th]
\psfig{figure=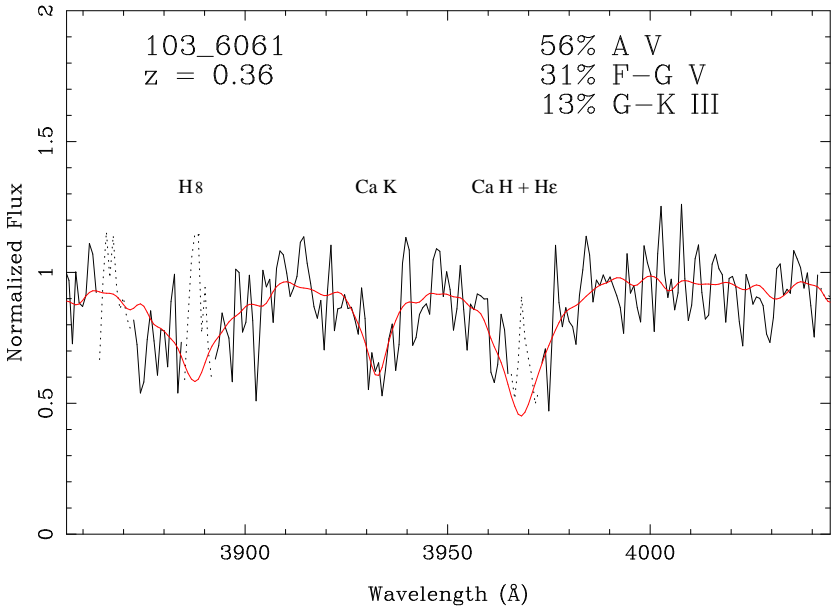,width=11.cm,angle=270}
\figcaption[im.fig8.ps]{
  The absorption-line spectrum of 103\_6061 compared with the best-fit 
 model spectrum. The spectrum is redshifted to the rest-frame and
 divided by the mean local continuum.
  The absorption-line velocity width ($\sigma_{abs}$)
 is measured  using the maximum likelihood method as described in the text. 
  We find $\sigma_{abs} < 80$ km\,sec$^{-1}$.   
 The best-fit mix of stellar types is indicated in the figure.   
\label{fig:spec_abs}
}
\end{figure}

  A more direct test would be to compare both 
 $\sigma_{em}$ and $\sigma_{abs}$ of the underlying old stellar population of 
 BSCs. 
  Unfortunately, our BSCs are in general too faint to 
 provide continuum spectra with enough S/N for the measurement of 
 $\sigma_{abs}$. For this reason,
 we derive a limit on $\sigma_{abs}$ for 
 only one object (103\_6061) in our sample.
  Fig. \ref{fig:spec_abs} shows a continuum-divided,  
 absorption-line spectrum of 103\_6061, from which we 
 obtain an upper limit  $\sigma_{abs} < 80$ km\,sec$^{-1}$. 
  The observed spectrum is fitted 
 with a model spectrum using a maximum-penalized likelihood method
 (Gebhardt et al. 2001; also see Saha \& Williams 1994; Merritt 1997).  
  The model spectrum is comprised of high-resolution, stellar 
 templates which are convolved according to a 
 given velocity dispersion and instrument resolution and are mixed with
 proper weights to provide the best match to the galaxy spectrum. 
  For more details on the absorption line fitting method,  
 see Gebhardt et al. (2001). 
  Some absorption lines in this particular object are contaminated 
 by emission lines (dotted line in Fig. \ref{fig:spec_abs}), and 
 we artificially remove these lines when fitting the spectrum.
  In the figure, Balmer emission lines seem narrower than the stellar
 absorption lines, but this is simply because stellar lines 
 are intrinsically broader than the nebular emission lines 
 in the absence of velocity fields. 
  The upper limit on $\sigma_{abs}$ (80 km sec$^{-1}$) for 
 103\_6061 is perfectly consistent with its $\sigma_{em}$ value of
 61 km sec$^{-1}$, supporting  the assumption
 that $\sigma_{em}$ reflects the galaxy mass,  or at least that  
 $\sigma_{em}$ does not severely underestimate the kinematics of the galaxy.
 
  As an alternative check,  we use 
 the fundamental plane relation of field, early-type 
 galaxies at $z \lesssim 1$ (Gebhardt et al. 2001) 
 to estimate the velocity dispersion of the red,
 underlying components (hereafter, $\sigma_{FP}$). 
 This approach assumes that
 BSCs with underlying red, stellar populations, as defined in the previous 
 section, are early-type galaxies
 with a sprinkling of young stars at their center.
  The FP relation in Gebhardt et al. (2001) may be 
 written as, 

\begin{equation}
 log_{10}(\sigma_{FP}{\rm (km\,s^{-1})})
 = 0.8 ~[log_{10}(r_{hl,under}{\rm (kpc)})
 + 0.32 \langle SB_{hl,under} \rangle + 9.062)],
\end{equation}

\noindent
 where $r_{hl}$ has been replaced by $r_{hl,under}$,
 the circular aperture half-light radius of
 the underlying, red component. 
 $SB_{hl,under}$ is the rest-frame $B$-band 
 surface brightness within $r_{hl,under}$ of the same component  
 corrected for the luminosity 
 evolution derived in Gebhardt et al. (2001). 
  Since there is uncertainty regarding the plausible surface brightness
 profile for BSCs, we use structural parameters from both 
 the simultaneous $r^{1/4}$+disk fit (Table 2) 
 and the double exponential fit (Table 3), 
 and use them to derive independent $\sigma_{FP}$ values.
 The quantity $r_{hl,under}$ is derived as 
 $r_{hl,under}=\sqrt{1-e_{b}} \times r_{e, \rm maj}$ from Table 2 
 when the underlying red component is fitted with a $r^{1/4}$ law. 
  When the underlying red component is fitted with an exponential 
 profile, we use 
 $r_{hl,under}=1.68 \, \sqrt{cos(inc)} \, r_{d,maj}$ from Table 2 or 
 Table 3.

  In Table 4, we compare these $\sigma_{FP}$ 
 versus $\sigma_{em}$.
  Note that the numbers within parentheses are the values derived
 from a double exponential profile.  
   First, we find that $\sigma_{FP}$ using the $r^{1/4}$+disk fit differ
 from $\sigma_{FP}$ derived using the double exponential fit.
  In many cases, the difference between two $\sigma_{FP}$ values is 
 small ($\pm$10 -- 20\%), but in some cases, $\sigma_{FP}$ values from
 these different methods differ almost by a factor of two
 (172\_5049 and 103\_6061).

\begin{figure}[ht]
\psfig{figure=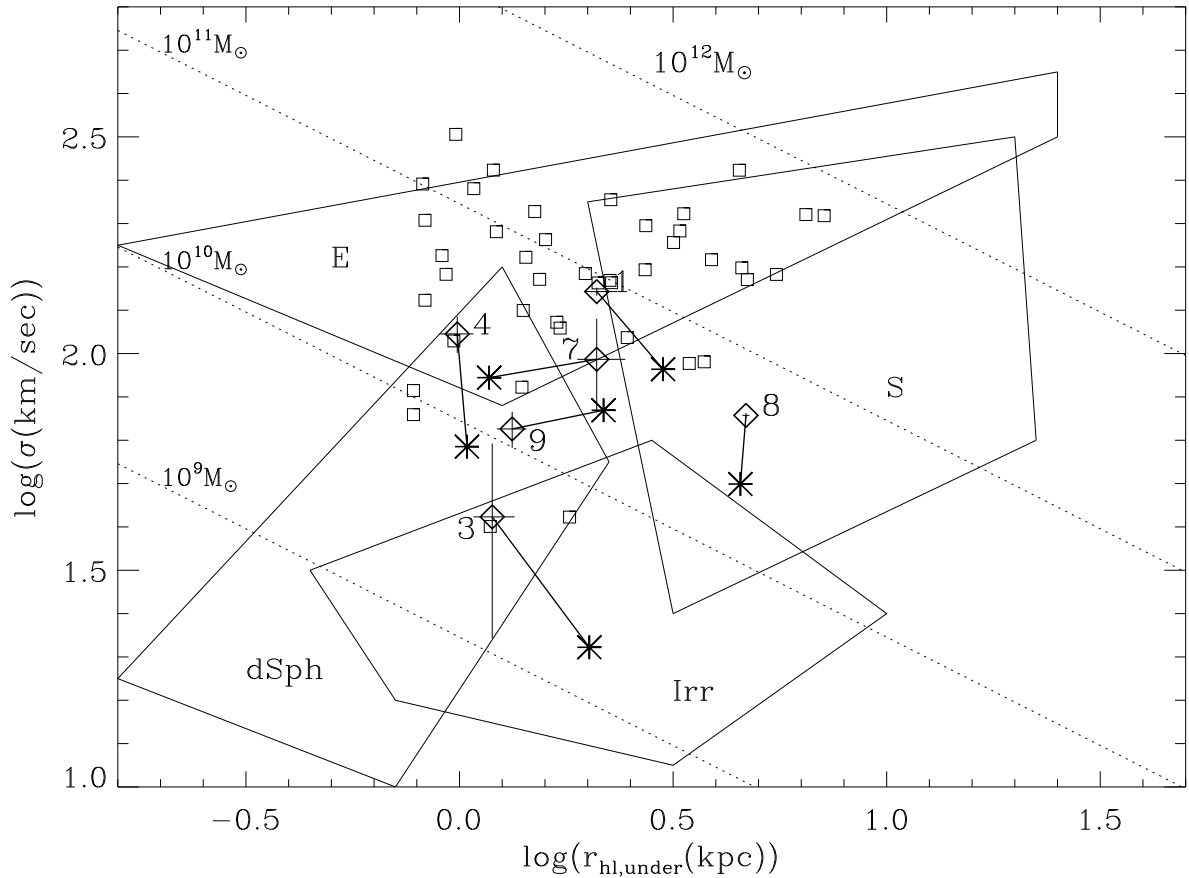,width=17.3cm,angle=90}
\figcaption[im.fig9.ps]{ Distribution of 
 $\sigma_{FP}$ vs. $R_{hl,under}$ for the underlying red 
 components of Type-1 BSCs.
  The spaces occupied by typical local Hubble types are taken 
 from Phillips et al. (1997). Big diamonds indicate 
 results based on a $r^{1/4}$ photo-bulge + exponential disk fit, while 
 asterisks are for a double exponential fit. The line connecting 
 diamonds and asterisks represents for each BSC 
 how the position changes when the fitting method changes.
  The number plotted next to each point is the object number that 
 appears in the first column of Table 1. The dotted lines show loci 
 of constant mass from Eq. (1). 
  Red, QS-E/S0 galaxies with spectroscopic redshifts in Im et al. (2001)
 are plotted with squares as in Fig. 5.
  Even using parameters of the underlying stellar population,
  BSCs appear to be low-mass objects. 
\label{fig:resig_fp}
}
\end{figure}

  Second, the comparison between the $\sigma$ values shows
 that $\sigma_{em}$ values 
 are similar to or smaller than $\sigma_{FP}$ by $20 \pm 20$\% on 
 average.
  The derived ratio, $\frac{\sigma_{em}}{\sigma_{FP}}$, is consistent with
 similar values derived from the comparison of velocity widths measured in
 HI, optical emission lines, and optical absorption lines for local 
 galaxies (Kobulnicky \& Gebhardt 2000; Telles \& Terlevich 1993; Rix et al.
  1997). The amount of discrepancy is small enough to ensure that mass
 of BSCs in Table 1 should not be underestimated by more than a factor of
 $\sim$2.
  In Fig. \ref{fig:resig_fp},
 we show $r_{hl,under}$ vs. $\sigma_{FP}$ of BSCs in Table 4.
  Like Fig. \ref{fig:resig}, we also plot 
 the same relation for distant, red, QS-E/S0s (squares). 
  The diamonds show $r_{hl,under}$ vs $\sigma_{FP}$ for 
 the $r^{1/4}$ + disk fit, and the stars indicate the same 
 relation for the double exponential fit.  
  Fig. \ref{fig:resig} shows that 
 BSCs still lie in the region which is populated by low-mass Es and dEs
 ($M_{dyn} < $ a few $\times 10^{10}\,M_{\odot}$) 
 even though we use $\sigma_{FP}$.
  A possible exception is 092\_1339 (\#1), which may have 
 a dynamical mass of $\sim 10^{11}\,M_{\odot}$. However, even this object 
 does not seem to be a typical massive E/S0 with $> 10^{11}\,M_{\odot}$. 

   In summary, 
   the $\sigma_{em}$ or ``emission line widths'' may not be a precise
 indicator of the kinematics of BSCs, but they seem to be useful as
 a rough estimator  of the 
 dynamical mass of BSCs (within a factor of $\sim$2 uncertainty).
   All the different sources of evidence presented here
 ($\sigma_{em}$ vs $\sigma_{abs}$ of 
 Es and a BSC, and $\sigma_{em}$ vs $\sigma_{FP}$ of BSCs)
 support our conclusion in section 3.1 that 
 Type-1 and Type-2 BSCs in our sample 
 are likely to be either progenitors of dwarf galaxies, or of low- or at most 
 moderate-mass E/S0s with $M_{dyn} \lesssim$
 a few $\times\, 10^{10}\,M_{\odot}$.
   However, this conclusion is based on relatively
 small number of objects, and more extensive studies are desired 
 to establish the link between $\sigma_{em}$ and $M_{dyn}$.  

\subsection{Comparison with other studies}

   Koo et al. (2001) study 68 high redshift ($z \sim 0.8$) 
 luminous photo-bulges ($I_{bulge} < 23.5$)
 in the Groth Strip; five of their objects 
 overlap with ours.
  They find  some ``blue'' photo-bulges, but these generally 
 have relatively low luminosities.
   Since they will fade  after the star-forming phase (e.g., Guzm\'an 
 et al.  1996), such blue photo-bulges will not evolve into
 the more luminous bulges today. 
 Their result is  consistent with our findings that BSCs are relatively
 low mass objects.

  In order to check for a possible link between CGs and BSCs, 
  we have also looked at values of $B/T$, $R$, and $V-I$ for CGs  
 in Phillips et al. (1997). Among 25 CGs at $I < 22$ with 
 known $z$, we find only 2 that qualify as BSCs. 
  The remaining 23 CGs turn out to be normal, red E/S0s (10),
 smooth galaxies with exponential profiles (6),
 or galaxies with disturbed morphologies (7).    

   Schade et al. (1999) studied elliptical galaxies in a subset of 
 the CFRS and LDSS samples for which HST images are available.  
 Eight ellipticals in the Schade et al. sample exist  
 in the Groth Strip.
  Among these, 5 objects we classify as red QS-E/S0s, and 
  one object is in our BSC sample (092\_1339). The
 remaining two objects are not classified as E/S0s by us
 (084\_1138 and 093\_3251). A visual inspection of these two galaxies 
 shows that they have red bulges surrounded by lumpy, blue low surface 
 brightness regions, suggesting that they may instead be spiral galaxies,
 or red, early-type galaxies which are accreting gaseous 
 blue satellites. All three possible blue ellipticals in Schade et al. (1999)
 mentioned above are either low mass, star-forming systems or galaxies with 
 non-smooth morphology.  
   On the other hand, one ``possible'' BSC in our sample, 062\_6465,
 is not classified as an elliptical in Schade et al. (1999).
  We suspect that this object is classified as S0 in their work,
 since its bulge fraction is $B/T \simeq 0.5$ and they classify objects
 with $B/T \sim 0.5$ as S0s.
   Also, note that Brinchmann et al. (1998) visually  
 classify this object as an E/S0.

  Menanteau et al. (1999; 2000) studied spheroidal galaxies (E or S0) 
 with HST and find that about 25\% of these spheroidal galaxies
 have blue colors.   More interestingly, they find that these 
 galaxies have a blue localized region and a red outer region
 similar to Type-1 BSCs.  
  Thus, their spheroids seem consistent with ours.
  A subset of their sample comes from Abraham et al. (1999),
 and in order to understand similarities and differences, 
 we have made a comparison of the Abraham et al. (1999) sample vs.  
 BSCs, which is given in the next subsection.

\subsection{BSCs in the Hubble Deep Field}

 The Hubble Deep Field (HDF) offers a much deeper HST WFPC2 image
 than our Groth Strip images.
  We have examined the HDF images for galaxies that fit 
 our selection criteria, and find seven QS-E/S0s with $I < 22$.
  Of these, only one (hd4-241.1) is a BSC. 
  In contrast, Bouwens et al. (1998) identify 11 E/S0s ($I < 22$) 
 in the HDF. Of these, Abraham et al. (1999) identify 4 with 
 signs of recent star formation 
 (see Menanteau et al. 2000 for  additional work on this subject).
  The BSC hd4-241.1 ($z=0.321$) is also found as
 a blue E/S0 by Abraham et al. (1999), but their 3 other blue E/S0s 
 are not identified as BSCs or QS-E/S0s in our study, due to 
 $R$ values which slightly exceed 0.1. 

  Three of their 4 blue E/S0s, including 
 hd4-241.1, have apparent magnitudes and sizes comparable to our BSCs,
 suggesting they are probably not massive E/S0s.
   The remaining object in their list (hd2-251.0 at $z\sim1$) is somewhat
 exceptional, in the sense that it is a bright object ($I \simeq 20$) 
 at high redshift for which we do not find a similar counterpart
 in our BSC sample. This object is also quite red and has a blue core.   
  However, we note that this object has 
 a very broad MgII emission line indicating the existence of an AGN
 (Phillips et al. 1997), and
 is also detected as a radio (Fomalont et al. 1997) and X-ray source 
 (Hornschemeier et al. 2000). Thus, the blue 
 nuclear color of this object probably originates from the AGN.
  On the other hand, the $\sigma$ measurement from absorption lines 
 gives $\sigma \simeq 195$ km/sec (Gebhardt et al. 2001); 
 thus the underlying old stellar population
 seems to be a massive system. 
  We did not classify this object as an QS-E/S0
 due to its slightly disturbed appearance
 ($R=0.13$) showing circular ripples of the type that can occur
 in merger remnants (e.g., Hernquist \& Quinn 1987).
  In summary, our BSC criteria demand greater smoothness and
 symmetry than Menanteau et al. (1999, 2000) or Abraham et al. (1999),
 who find roughly twice as many blue spheroids in total.  However,
 their blue E/S0s seem otherwise to have similar 
 structural and photometric properties to our BSCs.  Many even have
 blue localized regions and red envelopes like our Type-1 BSCs.

\subsection{Number density of blue massive early-type galaxies}
 
  We find that $\sim$13 --- 20\% of QS-E/S0s with $I < 22$ 
 are BSCs, depending on selection criteria.
  Among them, we estimate that none or perhaps only one (092\_1339) 
 out of 10 BSCs {\it may} become a massive early-type galaxy today with 
 $\sigma \gtrsim 150$ km/sec or $M_{dyn} \gtrsim 10^{11}\,M_{\odot}$.
  Thus, the fraction of truly massive blue E/S0s is found to be 
 at most a few percent or less of the QS-E/S0 sample.  
  This fraction might be larger since 
 a star forming massive early-type galaxy   
 shortly after merging may appear peculiar, thus falling outside  
 our selection criteria.  Studies of red, field E/S0s show that their number 
 density at $z \lesssim 1$ is not much different from that of
 present-day E/S0s, providing an independent constraint on the
 fraction of blue merger remnants: the current uncertainty on
 the number density estimate  for red, field E/S0s at $z \lesssim 1$ is
 about 30\%, therefore the possible fraction of blue merger remnants is
 $\lesssim 30$\% of E/S0s assuming that merger events tend to  
 create E/S0s than destroying/transforming them
 (e.g., Im et al. 2001,1999, 1996; Driver et al. 1998).

\subsection{Close neighbors}

  Barton et al. (2000) studied close pairs of local galaxies
 (separation, $\Delta r < 50$ kpc),
 and find that star forming, bulge-like systems are 
 possibly accreting gas from their close neighbors.
  By searching for close neighbors, we can test this scenario.
  The images of BSCs in Fig. \ref{fig:besoimage} have  a 
 physical scale of 50 kpc by 50 kpc. Four BSCs have 
 companion galaxies with similar apparent magnitudes
 (092\_1339, 103\_6061, 172\_5049, and 212\_1030). Redshifts of 
 all these companions are available \footnote{The redshifts of companions are
 $z_{comp}=1.55$ for 092\_1339 ($z=0.90$),
 $z_{comp}=0.65$ for 103\_6061 ($z=0.36$),
 $z_{comp}=1.0290$ for 172\_5049 ($z=0.36$),
 and $z_{comp}=0.88$ (left) and $z_{comp}=0.24$ (right) for 212\_1030
 ($z=0.88$).}, 
 which show that the apparent companions of three of the BSCs are not
 physically associated with the BSCs. The   
 only BSC  with a physically associated companion  
 within $\Delta r < 25$ kpc is 212\_1030, which is probably not a genuine BSC
 as we argued in the section 3.1. 
  Thus, if the blue colors of BSCs are due to accreted gas,
 the entire companion was swallowed, leaving no discernible remnants.

\section{Conclusion}

  From our DEEP GSS data containing 262 galaxies with $I < 22$,
 we have identified 10 blue spheroid 
 candidates at $z \lesssim 1$, which we define as 
 blue, bulge-dominated ($B/T > 0.4$), smooth galaxies 
 morphologically similar to local E/S0 galaxies. 
  We find that their internal velocity widths as 
 measured from optical emission lines are small,   
 $\sigma_{em} \lesssim 80$ km sec$^{-1}$.  Given their photometric, structural
 and kinematic properties, most BSCs  
 resemble local H II region-like galaxies (e.g., Phillips et al. 1986),
 and appear to be the progenitors of low-mass spheroidal galaxies today
 ($M_{dyn} \lesssim 10^{10}$ a few$\, M_{\odot}$).
  Two BSCs appear to be misclassified spiral galaxies, 
 and one BSC seems to be a red early-type galaxy
 with blue close neighbors.
  We identify possible underlying red stellar components in 5 or 6 out of
  10 BSCs (``Type-1''). Velocity dispersions 
 of the underlying red stellar component estimated from the FP 
 relation or from a direct measurement of the absorption-line 
 spectrum do not differ from $\sigma_{em}$ values
 by more than a factor of 1.2 on average.
  This implies that even though the $\sigma_{em}$ values may underestimate
 the true kinematics of the galaxy,
 the difference is not enough to affect the conclusion that 
 the underlying red components are not very massive
 ($\lesssim$ a few $\times\,10^{10} \, M_{\sun}$). 
  Overall,  we find that massive blue early-type galaxies with smooth,
 symmetric appearance are rare (a few percent of E/S0s with $I < 22$),
 suggesting either that the major merging events,
 which would produce star-forming
 massive, smooth, early-type galaxies, are rare at $z < 1$,
 or that the merger-triggered
 star formation activities subside by the time the merger products  
 appear smooth and symmetric.  Given our small sample size, 
 we encourage future studies of more BSCs, to obtain a more conclusive result 
 and to find any trend in the abundance and mass of BSCs as a function of
 redshift.

\acknowledgements
 This paper is based on observations with the NASA/ESA Hubble Space Telescope,
 obtained at the Space Telescope Science Institute, which is operated by
 the Association of Universities for Research in Astronomy, Inc., under
 NASA contract NAS5-26555.
 Funding for DEEP was provided by NSF grant AST-9529098.  
 This work was also supported by the STScI grants GO-07895.02-96A,
 AR-.6402.01-95A, and AR-07532.01-96. 
  We are grateful to Raja Guhathakurta,
 Garth Illingworth, Nicole Vogt, Vicki Sarajedini, and Andy Sheinis 
 for their help in acquisition and reduction of spectra,
 and useful discussions. We also thank the anonymous referee for useful
 comments and suggestions.


\clearpage

\vspace{2cm}

\clearpage

\begin{deluxetable}{c c  c c  c  c c c c c c c c c c l}
\tablefontsize{\scriptsize}
\tablecolumns{16}
\tablecaption{Basic Data of Blue Spheroid Candidates}
\tablewidth{0pt}
\tablehead{\# &  ID & RA & DEC & $z$ & $I_{tot}$ & $(V-I)$ & 
 $(U-B)_{rest}$ & $M_{B}$ &
 $B/T$  & $R$ & $\sigma_{em}$ &  $r_{hl,cir}$  & $M_{vir}$ & Type & 
 Remarks
 \\
 (1)  &  (2) &  (3) &  (4) &  (5) & (6) & (7) & (8) & (9) & (10) & (11) & (12) & (13) & (14) & (15) & } 
\startdata
 1 & $092\_1339$  & 14:17:27.411 & 52:26:44.58  &
             0.9031 & 21.40 & 1.30 & -0.10  & -21.22   & 0.87 &  0.05 & 
             $78$  & 2.2 & 1.9  &  1  &              \\ 
   & &         &    &    & $^{+0.03}_{-0.02}$  & $^{+0.04}_{-0.04}$ 
            & $^{+0.02}_{-0.03}$ & $^{+0.03}_{-0.03}$ & $^{+0.06}_{-0.05}$ &
            & $\pm 6$   & $^{+0.4}_{-0.3}$  &   &    &   \\     
 2 & $092\_4957$  & 14:17:23.163 & 52:26:53.45  & 
            0.2105 & 20.52 & 0.81 & -0.13  & -18.03   & 0.41 &  0.03 &  
             $61$ & 5.6 & 3.7 &     &  Rot. curve with \\ 
   & &        &     &    & $^{+0.03}_{-0.01}$  & $^{+0.03}_{-0.03}$ 
            & $^{+0.03}_{-0.03}$ & $^{+0.03}_{-0.03}$ & $^{+0.02}_{-0.03}$ &
            & $\pm 30$   & $^{+0.1}_{-0.1}$  &  &   & $V_{max}=130$km sec$^{-1}$ \\    
 3 & $103\_6061$  & 14:17:27.242  &  52:26:08.37  & 
             0.3631 & 21.44 & 0.91 & -0.15  & -18.47   & 0.49 &  0.04 &
             $61$ & 1.7 & 0.58  & 1  &                  \\ 
   & &        &    &     & $^{+0.03}_{-0.02}$  & $^{+0.02}_{-0.04}$ 
            & $^{+0.03}_{-0.03}$ & $^{+0.02}_{-0.02}$ & $^{+0.05}_{-0.06}$ &
            & $\pm 30$  & $^{+0.1}_{-0.1}$  &   &   &  \\     
 4 & $172\_5049$  &  14:16:31.410 & 52:17:26.24 &
             0.3564 & 21.18 & 1.07 & 0.03  & -18.51   & 0.73 &  0.03 &
             $59$ & 1.7 & 0.54 & 1  &                  \\ 
    & &       &    &     & $^{+0.03}_{-0.03}$  & $^{+0.04}_{-0.05}$ 
            & $^{+0.05}_{-0.05}$ & $^{+0.03}_{-0.05}$ & $^{+0.04}_{-0.05}$ &
            & $\pm 30$  & $^{+0.1}_{-0.1}$  &    &   &  \\     
 5 & $212\_1030$  & 14:16:10.228 & 52:12:37.11 & 
             0.8778 & 21.63 & 1.58 & 0.06  & -20.91   & 0.71 &  0.05 &
             72? & 3.8   & 2.8  &     &  Red early-type galaxy \\
   & &        &    &     & $^{+0.06}_{-0.04}$  & $^{+0.06}_{-0.08}$ 
            & $^{+0.04}_{-0.04}$ & $^{+0.06}_{-0.05}$ & $^{+0.14}_{-0.16}$ &
            & ...    & $^{+0.8}_{-1.1}$  &   &   & with a blue neighbor \\      
 6 & $294\_2078$  & 14:15:23.216 & 52:01:41.56 & 
            0.9295  & 21.95  & 1.44 & -0.01  & -20.81  & 0.59 &  0.00 & 
            $95$ & 4.8 & 6.3  &     &  Rot. curve; late-type?\\
   & &        &    &     & $^{+0.10}_{-0.06}$  & $^{+0.09}_{-0.13}$ 
            & $^{+0.06}_{-0.07}$ & $^{+0.12}_{-0.07}$ & $^{+0.07}_{-0.14}$ &
            & $\pm 30$ & $^{+0.6}_{-0.6}$  &  &   &           \\      
 7 & $062\_6465$\tablenotemark{*}  & 14:17:41.030 & 52:30:26.60 & 
            1.0196 & 21.89  & 1.69 & 0.17  & -21.35  & 0.51 &  0.08 &
             $29$ & 2.4 & 0.29  & 1  &                        \\ 
   & &       &     &    & $^{+0.05}_{-0.04}$  & $^{+0.06}_{-0.06}$ 
            & $^{+0.03}_{-0.03}$ & $^{+0.06}_{-0.05}$ & $^{+0.07}_{-0.07}$ &
            & $\pm 30$ & $^{+0.2}_{-0.2}$  &  &  &   \\     
 8 & $082\_5252$\tablenotemark{*}  & 14:17:29.473 & 52:27:58.56 & 
            0.3480  & 19.38  & 0.88 & -0.17  & -20.44  & 0.49 &  0.09 &
             ... & 4.9 & ...  &  1  &  $z$ from CFRS    \\
   & &        &     &    & $^{+0.01}_{-0.01}$  & $^{+0.01}_{-0.01}$ 
            & $^{+0.01}_{-0.01}$ & $^{+0.01}_{-0.01}$ & $^{+0.01}_{-0.01}$ &
            &     & $^{+0.1}_{-0.1}$  &   &   &         \\      
 9 & $153\_6056$\tablenotemark{*}  & 14:16:54.378 & 52:20:17.99 & 
             0.7206  & 21.63 & 1.14 & -0.22  & -20.19  & 0.36 &  0.05 &
             $79$ & 2.2 & 1.8 & 1 or 2  &               \\ 
   & &        &     &    & $^{+0.03}_{-0.03}$  & $^{+0.05}_{-0.06}$ 
            & $^{+0.04}_{-0.04}$ & $^{+0.03}_{-0.03}$ & $^{+0.03}_{-0.04}$ &
            & $\pm 30$  & $^{+0.1}_{-0.1}$  &  &  &   \\     
10 & $273\_4427$\tablenotemark{*}  &  14:15:34.140  & 52:05:53.96 & 
              0.9890  & 21.56 & 0.89 & -0.29  & -21.32  & 0.40 &  0.09 &  
             $86$ & 1.1 & 1.1  &  2   &              \\
   & &        &    &     & $^{+0.03}_{-0.03}$  & $^{+0.03}_{-0.04}$ 
            & $^{+0.02}_{-0.02}$ & $^{+0.04}_{-0.04}$ & $^{+0.03}_{-0.03}$ &
            & $\pm 30$  & $^{+0.1}_{-0.1}$  &  &  &           \\ 
\enddata
\tablecomments{(1) Object number as used in Fig. 5;
 (2) Source ID given by FFC-XXYY, where FF is the subfield,
 C is the WFPC2 chip number, and XX and YY are the chip coordinates in 
 units of 10 pixels;
(3) Right Ascension (J2000); (4) Declination (J2000); 
(5) Spectroscopic redshift;
(6) $I$-band total magnitude;
(7) Observed $V-I$ color derived from the total model-fit magnitudes; 
(8) Rest-frame $U-B$ color, which is derived from $V-I$ and $z$ using
 the $K$-correction in Gebhardt et al. (2001); 
(9) Rest-frame $B$-band absolute magnitude; 
(10) Bulge to total light ratio measured in $I$ band; 
(11) Residual parameter measured in $I$ band; 
(12) Velocity dispersion $\sigma_{em}$ measured from emission lines
 in km\,sec$^{-1}$; 
(13) Circular aperture half-light radius in pixels. One pixel 
corresponds to 0.1 arcsec; 
(14) Dynamical mass in units of $10^{10} M_{\sun}$.
 These mass estimates are uncertain by a factor of 2; 
(15) BSC Type as described in the text; 
 Errors associated with each quantity are $\pm 1 \sigma$.}
\tablenotetext{*}{``Possible'' BSCs. All others are ``good'' BSCs.}
\end{deluxetable}

\begin{deluxetable}{c c c c  c c c  c c c c c}
\tablefontsize{\scriptsize}
\tablecolumns{12}
\tablecaption{Structural Parameters of BSCs from Simultaneous $r^{1/4}$ + DISK Fit}
\tablewidth{0pt}
\tablehead{ \# & ID & $B/T$ & $r_{e,\rm maj}$\tablenotemark{a} &
 $r_{d, \rm maj}$\tablenotemark{b} & 
 $(V-I)_{b}$ & $(V-I)_{d}$ & $(U-B)_{b,rest}$ & $(U-B)_{d,rest}$ &
 $e_{b}$  & Inc & $\chi^{2}$ \\
 (1) & (2) & (3) & (4) & (5) & (6) & (7) & (8) & (9) & (10) & (11) & (12)} 
\startdata
1 & 092\_1339  &  0.85 &   3.4\tablenotemark{*}  &   0.6 &  1.63 &  0.45 &  0.10 & -0.64 &  0.25 &     9 &   1.030 \\
  &   & $^{+  0.03}_{-  0.04}$  & $^{+   0.4}_{-   0.2}$  & $^{+   0.1}_{-   0.1}$  & $^{+  0.08}_{-  0.07}$  & $^{+  0.26}_{-  0.30}$  & $^{+  0.05}_{-  0.04}$  & $^{+  0.17}_{-  0.20}$  & $^{+  0.02}_{-  0.02}$  & $^{+  2}_{-  6}$  &  \\
2 & 092\_4957  &  0.44 &  10.7 &   3.8 &  1.37 &  0.66 &  0.54 & -0.30 &  0.36 &    53 &   1.010 \\
  &   & $^{+  0.02}_{-  0.02}$  & $^{+   0.2}_{-   0.2}$  & $^{+   0.1}_{-   0.1}$  & $^{+  0.13}_{-  0.11}$  & $^{+  0.06}_{-  0.07}$  & $^{+  0.16}_{-  0.15}$  & $^{+  0.06}_{-  0.06}$  & $^{+  0.05}_{-  0.03}$  & $^{+  1}_{-  1}$  &  \\
3 & 103\_6061  &  0.50 &   3.1\tablenotemark{*} &   0.8 &  1.29 &  0.60 &  0.26 & -0.46 &  0.33 &    12 &   1.062 \\
  &   & $^{+  0.04}_{-  0.04}$  & $^{+   0.4}_{-   0.3}$  & $^{+   0.1}_{-   0.1}$  & $^{+  0.13}_{-  0.14}$  & $^{+  0.10}_{-  0.12}$  & $^{+  0.15}_{-  0.15}$  & $^{+  0.10}_{-  0.11}$  & $^{+  0.05}_{-  0.15}$  & $^{+ 17}_{-  8}$  &  \\
4 & 172\_5049  &  0.70 &   1.6 &   1.3\tablenotemark{*} &  0.78 &  2.16 & -0.28 &  1.39 &  0.27 &    17 &   1.024 \\
  &   & $^{+  0.03}_{-  0.02}$  & $^{+   0.1}_{-   0.1}$  & $^{+   0.1}_{-   0.1}$  & $^{+  0.05}_{-  0.05}$  & $^{+  0.26}_{-  0.24}$  & $^{+  0.05}_{-  0.05}$  & $^{+  0.37}_{-  0.33}$  & $^{+  0.02}_{-  0.02}$  & $^{+ 11}_{-  5}$  &  \\
5 & 212\_1030  &  0.52 &   1.6 &   5.4 &  1.79 &  1.59 &  0.19 &  0.08 &  0.02 &     5 &   1.051 \\
  &   & $^{+  0.04}_{-  0.05}$  & $^{+   0.2}_{-   0.2}$  & $^{+   0.3}_{-   0.3}$  & $^{+  0.20}_{-  0.19}$  & $^{+  0.20}_{-  0.15}$  & $^{+  0.12}_{-  0.11}$  & $^{+  0.12}_{-  0.09}$  & $^{+  0.02}_{-  0.02}$  & $^{+  5}_{-  4}$  &  \\
6 & 294\_2078  &  0.46 &   4.1 &   3.6 &  3.11 &  0.76 &  0.85 & -0.42 &  0.60 &    54 &   0.968 \\
  &   & $^{+  0.11}_{-  0.10}$  & $^{+   1.2}_{-   1.0}$  & $^{+   0.2}_{-   0.2}$  & $^{+  0.77}_{-  0.63}$  & $^{+  0.22}_{-  0.25}$  & $^{+  0.32}_{-  0.29}$  & $^{+  0.14}_{-  0.16}$  & $^{+  0.08}_{-  0.08}$  & $^{+  3}_{-  4}$  &  \\
7 & 062\_6465  &  0.51 &   3.0\tablenotemark{*} &   1.5 &  3.59 &  1.01 &  0.88 & -0.20 &  0.10 &    42 &   1.008 \\
  &   & $^{+  0.07}_{-  0.05}$  & $^{+   0.5}_{-   0.3}$  & $^{+   0.1}_{-   0.1}$  & $^{+  1.40}_{-  0.47}$  & $^{+  0.12}_{-  0.18}$  & $^{+  0.17}_{-  0.12}$  & $^{+  0.07}_{-  0.11}$  & $^{+  0.04}_{-  0.05}$  & $^{+  2}_{-  3}$  &  \\
8 & 082\_5252  &  0.47 &  10.6\tablenotemark{*} &   2.1 &  1.17 &  0.67 &  0.14 & -0.38 &  0.07 &     2 &   1.102 \\
  &   & $^{+  0.01}_{-  0.01}$  & $^{+   0.2}_{-   0.2}$  & $^{+   0.1}_{-   0.1}$  & $^{+  0.04}_{-  0.05}$  & $^{+  0.03}_{-  0.04}$  & $^{+  0.05}_{-  0.05}$  & $^{+  0.03}_{-  0.03}$  & $^{+  0.01}_{-  0.02}$  & $^{+  1}_{-  1}$  &  \\
9 & 153\_6056  &  0.33 &   2.9 &   1.5\tablenotemark{*} &  0.40 &  1.74 & -0.76 &  0.24 &  0.44 &    50 &   1.042 \\
  &   & $^{+  0.04}_{-  0.05}$  & $^{+   0.2}_{-   0.2}$  & $^{+   0.1}_{-   0.1}$  & $^{+  0.16}_{-  0.18}$  & $^{+  0.15}_{-  0.15}$  & $^{+  0.12}_{-  0.13}$  & $^{+  0.12}_{-  0.12}$  & $^{+  0.13}_{-  0.05}$  & $^{+  2}_{-  4}$  &  \\
10 & 273\_4427  &  0.39 &   3.3 &   0.7 &  1.01 &  0.79 & -0.22 & -0.36 &  0.41 &    74 &   1.083 \\
  &   & $^{+  0.04}_{-  0.03}$  & $^{+   0.7}_{-   0.9}$  & $^{+   0.1}_{-   0.1}$  & $^{+  0.19}_{-  0.14}$  & $^{+  0.09}_{-  0.11}$  & $^{+  0.11}_{-  0.09}$  & $^{+  0.06}_{-  0.07}$  & $^{+  0.08}_{-  0.24}$  & $^{+  1}_{-  1}$  &  \\
\enddata
\tablecomments{(1) Object number; (2) Object ID;
 (3) Bulge-to-total light ratio measured in the 
 $I$-band from the simultaneous fit; (4) Major axis bulge effective 
 radius in pixels. One pixel is equal to 0.1 arcsec; (5) Major axis 
 disk scale length in pixels; (6) Observed $V-I$ color of bulge; 
 (7) Observed $V-I$ color of disk;
 (8) Rest-frame $U-B$ color of bulge;
 (9)  Rest-frame $U-B$ color of disk;
 (10) Ellipticity of the bulge component;
 (11) Disk inclination angle in degrees. For a face-on system, this value
 is equal to 0; (12) Reduced $\chi^{2}$ value of the 2-D fit.}
\tablenotetext{a}{The circular aperture half-light radius, $r_{hl,cir}$,
 can be obtained as $r_{hl,cir}=\sqrt{1-e_{b}} \times r_{e, \rm major}$.}
\tablenotetext{b}{Major axis disk half light radius 
 can be obtained as $r_{hl,d}=1.68 \times r_{d, \rm major}$,
 and the circular aperture disk half light radius is 
 $r_{hl,cir}=\sqrt{cos({\rm Inc})} \times r_{hl,d}$.}
\tablenotetext{*}{Components which represent the
 underlying, red, stellar population for Type-1 BSCs.}
\end{deluxetable}

\begin{deluxetable}{c c c c  c c c  c c c c c}
\tablefontsize{\scriptsize}
\tablecolumns{12}
\tablecaption{Structural Parameters of BSCs from Simultaneous Double Exponential Fit}
\tablewidth{0pt}
\tablehead{\# & ID & $B/T$ & $r_{d1, \rm maj}$\tablenotemark{a,b} &
 $r_{d2, \rm maj}$\tablenotemark{a,b} & 
 $(V-I)_{d1}$ & $(V-I)_{d2}$ & $(U-B)_{d1,rest}$ & $(U-B)_{d2,rest}$  &
 $e_{d1}$  & Inc\tablenotemark{c} & $\chi^{2}$ \\
 (1) & (2) & (3) & (4) & (5) & (6) & (7) & (8) & (9) & (10) & (11) & (12)} 
\startdata
1 & 092\_1339  &  0.52 &   2.9\tablenotemark{*} &   0.6 &  1.57 &  1.09 &  0.06 & -0.23 &  0.23 &    46 &  1.013 \\
  &   & $^{+  0.05}_{-  0.04}$  & $^{+   0.4}_{-   0.4}$  & $^{+   0.1}_{-   0.1}$  & $^{+  0.18}_{-  0.17}$  & $^{+  0.15}_{-  0.15}$  & $^{+  0.11}_{-  0.10}$  & $^{+  0.09}_{-  0.09}$  & $^{+  0.06}_{-  0.06}$  & $^{+    4}_{-    7}$  &    \\
2 & 092\_4957  &  0.51 &   6.4 &   2.4 &  0.52 &  1.08 & -0.43 &  0.17 &  0.41 &    51 &  1.003 \\
  &   & $^{+  0.04}_{-  0.05}$  & $^{+   0.1}_{-   0.4}$  & $^{+   0.1}_{-   0.2}$  & $^{+  0.12}_{-  0.12}$  & $^{+  0.15}_{-  0.16}$  & $^{+  0.12}_{-  0.11}$  & $^{+  0.19}_{-  0.18}$  & $^{+  0.03}_{-  0.03}$  & $^{+    2}_{-    2}$  &    \\
3 & 103\_6061  &  0.71 &   0.8 &   2.7\tablenotemark{*} &  0.75 &  1.29 & -0.31 &  0.26 &  0.19 &    26 &  1.046 \\
  &   & $^{+  0.07}_{-  0.10}$  & $^{+   0.1}_{-   0.1}$  & $^{+   1.1}_{-   0.5}$  & $^{+  0.17}_{-  0.16}$  & $^{+  0.44}_{-  0.53}$  & $^{+  0.17}_{-  0.16}$  & $^{+  0.53}_{-  0.56}$  & $^{+  0.03}_{-  0.05}$  & $^{+   10}_{-   16}$  &    \\
4 & 172\_5049  &  0.73 &   1.5\tablenotemark{*} &   0.2 &  1.19 &  0.79 &  0.16 & -0.26 &  0.21 &    30 &  1.033 \\
  &   & $^{+  0.02}_{-  0.04}$  & $^{+   0.1}_{-   0.1}$  & $^{+   0.1}_{-   0.1}$  & $^{+  0.08}_{-  0.07}$  & $^{+  0.17}_{-  0.14}$  & $^{+  0.09}_{-  0.08}$  & $^{+  0.17}_{-  0.14}$  & $^{+  0.03}_{-  0.03}$  & $^{+    6}_{-    9}$  &    \\
5 & 212\_1030  &  0.78 &   2.6 &   0.1 &  1.84 &  1.74 &  0.22 &  0.16 &  0.18 &    80 &  1.054 \\
  &   & $^{+  0.02}_{-  0.03}$  & $^{+   0.3}_{-   0.3}$  & $^{+   0.2}_{-   0.1}$  & $^{+  0.16}_{-  0.15}$  & $^{+  0.26}_{-  0.24}$  & $^{+  0.10}_{-  0.09}$  & $^{+  0.16}_{-  0.15}$  & $^{+  0.08}_{-  0.09}$  & $^{+    4}_{-   14}$  &    \\
6 & 294\_2078  &  0.55 &   2.2 &   5.2 &  1.97 &  0.61 &  0.30 & -0.52 &  0.66 &    45 &  0.985 \\
  &   & $^{+  0.12}_{-  0.15}$  & $^{+   0.3}_{-   0.4}$  & $^{+   1.0}_{-   1.2}$  & $^{+  0.50}_{-  0.36}$  & $^{+  0.39}_{-  0.32}$  & $^{+  0.26}_{-  0.20}$  & $^{+  0.25}_{-  0.21}$  & $^{+  0.04}_{-  0.07}$  & $^{+   10}_{-   17}$  &    \\
7 & 062\_6465  &  0.48 &   2.0 &   1.0\tablenotemark{*} &  1.17 &  1.93 & -0.11 &  0.29 &  0.19 &    49 &  1.016 \\
  &   & $^{+  0.21}_{-  0.27}$  & $^{+   0.7}_{-   0.5}$  & $^{+   0.1}_{-   0.2}$  & $^{+  0.66}_{-  0.62}$  & $^{+  0.80}_{-  0.88}$  & $^{+  0.35}_{-  0.38}$  & $^{+  0.34}_{-  0.48}$  & $^{+  0.08}_{-  0.10}$  & $^{+   11}_{-   10}$  &    \\
8 & 082\_5252  &  0.34 &   6.4\tablenotemark{*} &   1.8 &  0.81 &  0.80 & -0.24 & -0.25 &  0.14 &     1 &  1.101 \\
  &   & $^{+  0.02}_{-  0.01}$  & $^{+   0.1}_{-   0.2}$  & $^{+   0.1}_{-   0.1}$  & $^{+  0.06}_{-  0.06}$  & $^{+  0.03}_{-  0.04}$  & $^{+  0.06}_{-  0.06}$  & $^{+  0.04}_{-  0.04}$  & $^{+  0.02}_{-  0.02}$  & $^{+    3}_{-    1}$  &    \\
9 & 153\_6056  &  0.48 &   0.9 &   2.6\tablenotemark{*} &  1.00 &  1.44 & -0.33 &  0.01 &  0.36 &    55 &  1.040 \\
  &   & $^{+  0.08}_{-  0.21}$  & $^{+   0.1}_{-   0.3}$  & $^{+   0.6}_{-   0.6}$  & $^{+  0.46}_{-  0.50}$  & $^{+  0.45}_{-  0.68}$  & $^{+  0.35}_{-  0.37}$  & $^{+  0.35}_{-  0.51}$  & $^{+  0.08}_{-  0.07}$  & $^{+    5}_{-    9}$  &    \\
10 & 273\_4427  &  0.24 &   2.4 &   0.6 &  0.90 &  0.83 & -0.29 & -0.34 &  0.39 &    73 &  1.083 \\
  &   & $^{+  0.04}_{-  0.05}$  & $^{+   0.1}_{-   0.2}$  & $^{+   0.1}_{-   0.1}$  & $^{+  0.28}_{-  0.24}$  & $^{+  0.09}_{-  0.09}$  & $^{+  0.17}_{-  0.15}$  & $^{+  0.06}_{-  0.05}$  & $^{+  0.08}_{-  0.10}$  & $^{+    3}_{-    2}$  &    \\
\enddata
\tablecomments{(1) Object number; (2) Object ID;
 (3) Bulge-to-total light ratio measured in the 
 $I$-band from the simultaneous fit; (4) Major axis exponential 
 scale length for exponential photo-bulge in pixels.
 One pixel is equal to 0.1 arcsec;
 (5) Major axis disk scale length in pixels;
 (6) Observed $V-I$ color of exponential photo-bulge; 
 (7) Observed $V-I$ color of photo-disk;
 (8) Rest-frame $U-B$ color of exponential photo-bulge;
 (9) Rest-frame $U-B$ color of photo-disk;
 (10) Ellipticity of the exponential photo-bulge component;
 (11) Disk inclination angle in degrees. For a face-on system, this value
 is equal to 0; (12) Reduced $\chi^{2}$ value of the 2-D fit.}
\tablenotetext{a}{The circular aperture half-light radius, $r_{hl,cir}$,
 can be obtained as $r_{hl,cir}=\sqrt{1-e_{d1(2)}} \times
 r_{d1(2),\rm maj}$.}
\tablenotetext{b}{Major axis disk half light radius 
 can be obtained as $r_{hl,d}=1.68 \times r_{d,\rm maj}$.
 The circular aperture half-light radius is 
 $r_{hl,cir}=\sqrt{cos(Inc)} \times r_{hl,d}$.}
\tablenotetext{c}{$e_{d2} = 1-cos({\rm Inc})$.}
\tablenotetext{*}{Components which represent the
 underlying, red, stellar population for Type-1 BSCs.}
\end{deluxetable}

\begin{deluxetable}{c c c c}
\tablecolumns{4}
\tablecaption{Comparison between Velocity Dispersion Estimates}
\tablewidth{0pt}
\tablehead{\# & ID & $\sigma_{FP}$(km\,s$^{-1}$) & $\frac{\sigma_{em}}{\sigma_{FP}}$
\\  (1) &  (2)  &  (3) & (4)}
\startdata
1 & $092\_1339$  & $136 (92) \pm 10$  &  $0.57 (0.85) \pm 0.06$   \\
3 & $103\_6061$  & $ 42 (21) \pm 4$  &  $1.44 (2.87) \pm 0.73$   \\
4 & $172\_5049$  & $111 (61) \pm 8$  &  $0.53 (0.96) \pm 0.27$   \\
7 & $062\_6465$  & $ 97 (88) \pm 14$ &  $0.29 (0.33) \pm 0.31$   \\ 
8 & $082\_5252$  & $ 72 (50) \pm 3$  &    ...             \\
9 & $153\_6056$  & $ 67 (74) \pm 7$  &  $1.18 (1.06) \pm 0.46$   \\
  &              &                   &                    \\
  &            & $\langle \frac{\sigma_{em}}{\sigma_{FP}} \rangle$\tablenotemark{a} $~~=$ & $0.81 (1.22) \pm 0.24$
\enddata
\tablecomments{(1) Object number; (2) Object ID;
 (3) Estimated $\sigma$ from fundamental plane
 relation for the underlying red component from the $r^{1/4}$ + disk fit and 
 from the double exponential fit (numbers in parentheses);
 (4) Ratio of $\sigma$ from emission lines ($\sigma_{em}$)
 to $\sigma_{FP}$}
\tablenotetext{a}{Mean and square-rooted variance of $\frac{\sigma_{em}}{\sigma_{FP}}$}
\end{deluxetable}

\end{document}